# The $(1,0)\oplus(0,1)$ spinor description of the photon field and its preliminary applications

**Zhi-Yong Wang***

School of Optoelectronic Information, University of Electronic Science and Technology of China, Chengdu 610054, CHINA

*E-mail: zywang@uestc.edu.cn

Because spatio-temporal tensors are associated with the Lorentz group, whereas spinors are associated with its covering group SL(2, C), one can associate with every tensor a spinor (but not *vice versa*). In particular, the $(1,0)\oplus(0,1)$ representation of SL(2, C) can provide a six-component spinor equivalent to the electromagnetic field tensor. The chief aim of this work is to develop the $(1,0)\oplus(0,1)$ description for the electromagnetic field in the absence of sources, rigorously and systematically, which should be useful if we are to deal with those issues involving with single-photon states and the angular momentum of light, etc. Based on our formalism, the quantum theory and some symmetries of the photon field can be discussed in a new manner, and the spin-orbit interaction of photons can be described in a form that is closely analogous to that of the Dirac electron. Moreover, in terms of the $(1,0)\oplus(0,1)$ description, one can treat the photon field in curved spacetime via spin connection and the tetrad formalism, which is of great advantage to study the gravitational spin-orbit coupling of photons.

## 1. Introduction

As we know, in the four-dimensional (4D) spacetime, tensors are associated with the Lorentz group, while spinors are associated with the group SL(2, C), which is the covering group of the Lorentz group. As a result of this fact, spinors can be used to describe particles of any spin, whereas tensors can describe only the particles with integer spins. In other words, the group SL(2, C) yields all representations, including those with half-integral spins, whereas the Lorentz group yields only the representations with integral spins, and the correspondence between the two groups is a homomorphism rather than an isomorphism. In fact, spinors are considered to be more fundamental than tensors from both the mathematical and the physical points of view. As a consequence, one can associate with every tensor a spinor, but not vice versa: not all spinors correspond to tensors [1].

Historically, several formalisms have been used for higher spin fields, where the 2(2S +1) formalism gives the equations which are in some sense on an equal footing with the Dirac equation [2-4]. In particular, people have introduced several kinds of photon wave functions (or spinors) and reformulated Maxwell equations as the corresponding Dirac-like equations [5-30], which can be regarded as optical analogues of the Dirac equation. The interest to this issue has grown in recent years [31-48]. For example, the current interest in entanglement and its application to quantum information has rekindled the controversy surrounding the photon wave function [49]. However, in some of the previous literatures, the so-called "spinors" describing the electromagnetic field are just formally analogies with the Dirac spinor, rather than a real spinor, while some investigations on spinors ([1, 4, 50, 51]) are so formalized and abstract that they are inconvenient for practical applications. In other words, a rigorous and detailed spinor description of the electromagnetic field has still been absent. Therefore, it is necessary for us to develop a rigorous spinor description of the electromagnetic field in a systematical manner, which is based on three chief reasons:

1). As we know, the simplest covariant massless field for helicity $\pm 1$ has the Lorentz transformation type $(1,0)\oplus(0,1)$, and it is impossible to construct a vector field for massless particles of helicity $\pm 1$ [2, 3]. This implies that it is advantageous to study the spin-orbit interaction of the photon field based on the $(1,0)\oplus(0,1)$ description.

2). In Ref. [1] all spinors equivalent to tensors are taken as two-component forms, they are too complicated and abstract to be applicable for some practical issues. In our case, the $(1,0)\oplus(0,1)$ spinor equivalent to the electromagnetic field tensor is a six-component form. Our formalism is more convenient and straightforward, based on which our theory can be developed in a manner being closely parallel to that of the $(1/2,0)\oplus(0,1/2)$ Dirac field.

3). Nowadays, single-photon states and the angular momentum of light involve with a wide and diverse variety of issues (e.g., they play very important roles in quantum information and quantum computation) [47, 48, 52-60], for those our $(1,0)\oplus(0,1)$ spinor description of the photon field can provide a useful tool.

The main purpose of the present work is to lay the foundations for our next work. Nevertheless, in this paper a preliminary and heuristic study on the spin-orbit interaction of photons in an inhomogeneous medium will be presented. Moreover, to show one of the merits of the $(1,0)\oplus(0,1)$ description, we will for the first time treat the photon field in curved spacetime by means of spin connection and the tetrad formalism, and study the effect of gravitational spin-orbit coupling on the circular photon orbit in the Schwarzschild geometry.

BTW, let us present some heuristic comments on the quantum mechanics of single photons (in view of the fact that there is no unique vacuum sate in a general spacetime, all of our statements are presented from the point of view of an inertial observer):

As we know, in terms of generalized coordinates and generalized momentums, the quantized procedure of a field can be developed within the framework of the first quantization, such that the notion of the second quantization seems to be unwanted and can be replaced with that of the field quantization. In spite of that, for convenience let us distinguish between the quantum-mechanical behaviors in the sense of the first quantization and the quantum field-theoretical behaviors in the sense of the second quantization (though all of them are unified in quantum field theory). For example, the behaviors such as the quantum interferences and quantum tunneling of single particles, and the energy quantization of a particle in a potential well, etc., are called the quantum-mechanical behaviors of single particles, while the behaviors such as the creations and annihilations of particles, the vacuum fluctuation of a field, as well as the fact that a field consists of field quanta, etc, are called the quantum field-theoretical behaviors (they involve many-particle phenomena). According to quantum field theory, the laws of nature take the form of a quantum theory of fields, while all elementary particles are the quantum excitations of fields. On the other hand, each of field quanta, as a single particle, should be governed by the laws of quantum mechanics in the sense of the first quantization.

The photon is the only particle that was known as a field before it was detected as a particle, and the quantum theory of the electromagnetic field is directly a quantum field theory, such that the first-quantized quantum mechanics of single photons has been neglected, to a great extent. However, photons, as the field quanta of the electromagnetic field, should also obey the laws of single-particle's quantum mechanics. In other words, when there is not any many-particle phenomenon, one can deal with single-photon states which obey quantum mechanics in the sense of the first quantization. In fact, the two-slit interference phenomenon of *classical* electromagnetic waves can be attributed to the quantum mechanical behavior of single photons, and it is different from the quantum field-theoretical behavior such as the creations and annihilations of photons, the vacuum fluctuations, etc. The $(1,0)\oplus(0,1)$ spinor description of the classical electromagnetic field can provide the quantum mechanics of single photons in the sense of the first quantization, whose field quantization provides the quantum field theory of the photon field. It is unessential whether the $(1,0)\oplus(0,1)$ spinor can be regarded as a photon wave function or not, it is a field operator from the point of view of quantum field theory.

We work in the natural units, $\hbar = c = 1$, the 4D Minkowski metric tensor is taken as $\eta_{\mu\nu} = \mathrm{diag}(-1,1,1,1)$ ($\mu,\nu = 0,1,2,3$). Complex conjugation is denoted by $*$ and hermitian conjugation by $\dagger$. Repeated indices must be summed according to the Einstein rule.

## 2. The chiral and standard representations of the $(1,0)\oplus(0,1)$ spinor

People have introduced several photon wave functions, some of them are not a real spinor. Here we will describe the electromagnetic field in terms of a 6-component spinor that transforms according to the $(1,0)\oplus(0,1)$ representation of the group SL(2, C). To lay a soild and self-contained foundation for our future investigations, let us first discuss the chiral and standard representations of the $(1,0)\oplus(0,1)$ spinor, in the manner of discussing the ones of the $(1/2,0)\oplus(0,1/2)$ spinor (i.e., the Dirac spinor), by which we will present a general and unified result (e.g., the Riemann-Silberstein vector is a special case in our framework). In other words, we can consider the fields $(1/2,0)\oplus(0,1/2)$ and $(1,0)\oplus(0,1)$ on an equal footing, from the group theoretical viewpoint.

As we know, the most general (proper) Lorentz transformation is composed of rotations about three axes, and boosts in three directions, where the rotations generators $\boldsymbol{J}=(J_1,J_2,J_3)$ and the boost generators $\boldsymbol{K}=(K_1,K_2,K_3)$ satisfy the Lie algebras:

$$[J_l,J_m]=\mathrm{i}\varepsilon_{lmn}J_n\,,\ \ [K_l,K_m]=-\mathrm{i}\varepsilon_{lmn}J_n\,,\ \ [J_l,K_m]=\mathrm{i}\varepsilon_{lmn}K_n\,, \tag{1}$$

where $\varepsilon_{klm}=\varepsilon^{klm}$ denote the totally antisymmetric tensor with $\varepsilon_{123}=1$, $k,l,m=1,2,3$. Let

$$\boldsymbol{\Omega}^{(\pm)}=(\boldsymbol{J}\pm\mathrm{i}\boldsymbol{K})/2\,, \tag{2}$$

one has the Lie algebras:

$$[\Omega_l^{(+)},\Omega_m^{(+)}]=\mathrm{i}\varepsilon_{lmn}\Omega_n^{(+)}\,,\ \ [\Omega_l^{(-)},\Omega_m^{(-)}]=\mathrm{i}\varepsilon_{lmn}\Omega_n^{(-)}\,,\ \ [\Omega_l^{(+)},\Omega_m^{(-)}]=0\,. \tag{3}$$

Then $\boldsymbol{\Omega}^{(+)}$ and $\boldsymbol{\Omega}^{(-)}$ each generate a group SU(2), and the two groups commute. The Lorentz group is then essentially $\mathrm{SU}(2)\otimes\mathrm{SU}(2)$, and states transforming in a well-defined way will be labeled by two angular momenta $(j,j')$, the first one corresponding to $\boldsymbol{\Omega}^{(+)}$, and the second to $\boldsymbol{\Omega}^{(-)}$. As special cases, one or the other will correspond to spin zero (see for example, Refs. [61, 3]):

$$\boldsymbol{J}=\mathrm{i}\boldsymbol{K}\,,\ \ \boldsymbol{\Omega}^{(-)}=0\rightarrow(j,0)\,;\ \ \boldsymbol{J}=-\mathrm{i}\boldsymbol{K}\,,\ \ \boldsymbol{\Omega}^{(+)}=0\rightarrow(0,j)\,. \tag{4}$$

Under space inversion (parity operation), the boost generators $\boldsymbol{K}$ change sign ($\boldsymbol{K}$ is a vector), while the rotations generators $\boldsymbol{J}$ do not ($\boldsymbol{J}$ is a axial vector), such that the $(j,0)$ and $(0,j)$ representations become interchanged, $(j,0)\leftrightarrow(0,j)$.

Correspondingly, there are six parameters related to the three angles and three velocities of the general Lorentz transformations. The angle parameters in a 3D rotation about an axis $\boldsymbol{n}$ through an angle $\theta$ is denoted as $\boldsymbol{\theta} = \boldsymbol{n}\theta = (\theta_1, \theta_2, \theta_3)$, and the velocity parameters in a pure Lorentz transformation with relative velocity $\boldsymbol{v} = v\boldsymbol{n}'$ along an axis $\boldsymbol{n}'$ is denoted as $\varsigma = \boldsymbol{n}'\varsigma = (\varsigma_1, \varsigma_2, \varsigma_3)$, where $\varsigma = \sqrt{\varsigma_1^2 + \varsigma_2^2 + \varsigma_3^2}$ and $\tanh\varsigma = v/c$, $c$ is the velocity of light in Minkowski vacuum (we work in the natural units, $\hbar = c = 1$). The six parameters $\boldsymbol{\theta}$ and $\varsigma$ can form an antisymmetric tensor $\omega^{\mu\nu} = -\omega^{\nu\mu}$ with $\omega^{l0} = -\omega^{0l} = \varsigma^l$ and $\omega^{lm} = -\varepsilon^{lmn}\theta_n$. Under a Lorentz transformation ( $x^\mu \to x'^\mu = \Lambda^\mu{}_\nu x^\nu$, or $x \to x' = \Lambda x$ ) parametrized by $\omega^{\mu\nu}$, a field quantity $\psi$ is called a spinor provided that it transforms in the following manner:

$$\psi(x) \to \psi'(x') = \exp(-\mathrm{i}\omega_{\mu\nu} S^{\mu\nu}/2)\psi(x) = L(\Lambda)\psi(x), \tag{5}$$

where $L(\Lambda) = \exp(-\mathrm{i}\omega_{\mu\nu} S^{\mu\nu}/2)$ is called the spinor representation of the Lorentz transformation $\Lambda$, the antisymmetric tensor $S^{\mu\nu} = -S^{\nu\mu}$ is the 4D spin tensor of the field $\psi$, it can be formed by means of the rotations generators $\boldsymbol{J}$ and the boost generators $\boldsymbol{K}$. One can also refer to $\psi$ as the spinor representation of the Lorentz group.

Let $\partial_\mu = \partial/\partial x^\mu$ with $x^\mu = (t, \boldsymbol{x})$, $\nabla = (\partial_1, \partial_2, \partial_3)$, $\partial_t = \partial_0 = \partial/\partial t$. In Minkowski vacuum, the electromagnetic field intensities $\boldsymbol{E} = (E_1, E_2, E_3)$ and $\boldsymbol{H} = (H_1, H_2, H_3)$ satisfy the Maxwell equations

$$\nabla \times \boldsymbol{H} = \partial\boldsymbol{E}/\partial t, \quad \nabla \times \boldsymbol{E} = -\partial\boldsymbol{H}/\partial t, \tag{6-1}$$

$$\nabla \cdot \boldsymbol{E} = 0, \quad \nabla \cdot \boldsymbol{H} = 0. \tag{6-2}$$

Using the 4D Fourier transforms of $\boldsymbol{E}$ and $\boldsymbol{H}$ one can derive the transversality conditions (given by Eq. (6-2)) from the dynamical equations (given by Eq. (6-1)), such that the Maxwell equations in Minkowski vacuum can be represented only by Eq. (6-1). When we describe the electromagnetic field in terms of a 6-component spinor that transforms according to the $(1,0) \oplus (0,1)$ representation of the group SL(2, C), the 6-component spinor is called the $(1,0)\oplus(0,1)$ spinor. For the spinor we will introduce two representations, i.e., the chiral and standard representations, they are analogous to the chiral and standard representations of the $(1/2,0)\oplus(0,1/2)$ spinor (i.e., the Dirac spinor).

In this paper, *the column-matrix forms* of the vectors $\boldsymbol{E}$ and $\boldsymbol{H}$ are also denoted as $\boldsymbol{E}$ and $\boldsymbol{H}$, i.e., $\boldsymbol{E}=\begin{pmatrix}E_1 & E_2 & E_3\end{pmatrix}^{\mathrm{T}}$, $\boldsymbol{H}=\begin{pmatrix}H_1 & H_2 & H_3\end{pmatrix}^{\mathrm{T}}$ (the superscript T denotes the matrix transpose, the same below). In the chiral representation, the $(1,0)\oplus(0,1)$ spinor can be formed by means of the column-matrix form of $\boldsymbol{E}$ and $\boldsymbol{H}$:

$$\psi_{\mathrm{C}}=\frac{1}{2}\begin{pmatrix}\boldsymbol{E}+\mathrm{i}\boldsymbol{H}\\ \boldsymbol{E}-\mathrm{i}\boldsymbol{H}\end{pmatrix}=\begin{pmatrix}\boldsymbol{F}_{\mathrm{R}}\\ \boldsymbol{F}_{\mathrm{L}}\end{pmatrix}, \tag{7}$$

where

$$\boldsymbol{F}_{\mathrm{R}}=(\boldsymbol{E}+\mathrm{i}\boldsymbol{H})/2,\ \ \boldsymbol{F}_{\mathrm{L}}=(\boldsymbol{E}-\mathrm{i}\boldsymbol{H})/2. \tag{8}$$

In terms of $\psi_{\mathrm{C}}$, the Maxwell equations in Minkowski vacuum can be expressed as a single equation (the Dirac-like equation):

$$\mathrm{i}\beta_{\mathrm{C}}^{\mu}\partial_{\mu}\psi_{\mathrm{C}}(x)=0, \tag{9}$$

where $\beta_{\mathrm{C}}^{\mu}=(\beta_{\mathrm{C}}^{0},\boldsymbol{\beta}_{\mathrm{C}})$ satisfy $\beta_{\mathrm{C}}^{0}\boldsymbol{\beta}_{\mathrm{C}}=-\boldsymbol{\beta}_{\mathrm{C}}\beta_{\mathrm{C}}^{0}$, $\beta_{\mathrm{C}}^{0}=-\beta_{\mathrm{C}0}$, $\beta_{\mathrm{C}}^{l}=\beta_{\mathrm{C}l}$ ($l=1,2,3$), and, in the chiral representation, in $3\times3$ block form, they are ( $I_{n\times n}$ denotes the $n\times n$ unit matrix , $n=2,3,\ldots$),

$$\beta_{\mathrm{C}}^{0}=\begin{pmatrix}0 & I_{3\times3}\\ I_{3\times3} & 0\end{pmatrix},\ \ \boldsymbol{\beta}_{\mathrm{C}}=\begin{pmatrix}0 & -\boldsymbol{\tau}\\ \boldsymbol{\tau} & 0\end{pmatrix}, \tag{10}$$

where the matrix vector $\boldsymbol{\tau}=(\tau_1,\tau_2,\tau_3)$ consists of the following components:

$$\tau_1=\begin{pmatrix}0&0&0\\0&0&-\mathrm{i}\\0&\mathrm{i}&0\end{pmatrix},\ \ \tau_2=\begin{pmatrix}0&0&\mathrm{i}\\0&0&0\\-\mathrm{i}&0&0\end{pmatrix},\ \ \tau_3=\begin{pmatrix}0&-\mathrm{i}&0\\\mathrm{i}&0&0\\0&0&0\end{pmatrix}. \tag{11}$$

Moreover, in the chiral representation, one can also define the following matrices:

$$\boldsymbol{\alpha}_{\mathrm{C}}=\beta_{\mathrm{C}}^{0}\boldsymbol{\beta}_{\mathrm{C}}=\begin{pmatrix}\boldsymbol{\tau} & 0\\ 0 & -\boldsymbol{\tau}\end{pmatrix},\ \ \boldsymbol{\Sigma}_{\mathrm{C}}=\begin{pmatrix}\boldsymbol{\tau} & 0\\ 0 & \boldsymbol{\tau}\end{pmatrix},\ \ \beta_{\mathrm{C}}^{5}=\beta_{\mathrm{C}5}=\begin{pmatrix}I_{3\times3} & 0\\ 0 & -I_{3\times3}\end{pmatrix}. \tag{12}$$

One can show that $\beta_{\mathrm{C}}^{5}\beta_{\mathrm{C}}^{\mu}=-\beta_{\mathrm{C}}^{\mu}\beta_{\mathrm{C}}^{5}$. Eq. (9) can be rewritten as

$$\mathrm{i}\partial_{t}\psi_{\mathrm{C}}(x)=\hat{H}\psi_{\mathrm{C}}(x)\,, \tag{13}$$

where $\hat{H}=-\mathrm{i}\boldsymbol{\alpha}_{\mathrm{C}}\cdot\nabla$ represents the Hamiltonian of free photons in Minkowski vacuum. Similar to the Dirac theory of electron, let $\hat{\boldsymbol{L}}=\boldsymbol{x}\times(-\mathrm{i}\nabla)$ be the orbital angular momentum operator, one can easily prove $[\hat{H},\hat{\boldsymbol{L}}+\boldsymbol{\Sigma}_{\mathrm{C}}]=0$, where $\boldsymbol{\Sigma}_{\mathrm{C}}$ given by Eq. (12) represents the 3D spin matrix of photons and satisfies $\boldsymbol{\Sigma}_{\mathrm{C}}\cdot\boldsymbol{\Sigma}_{\mathrm{C}}=1(1+1)I_{6\times6}$. Moreover, let $\psi_{\mathrm{C}}{}^{\dagger}$ denote the Hermitian adjoint of $\psi_{\mathrm{C}}$ and $\bar{\psi}_{\mathrm{C}}=\psi_{\mathrm{C}}{}^{\dagger}\beta_{C}^{0}$, one can show that the quantity $\bar{\psi}_{\mathrm{C}}\psi_{\mathrm{C}}$ is a Lorentz scalar, while $\psi_{\mathrm{C}}{}^{\dagger}\psi_{\mathrm{C}}$ corresponds to the energy density of the photon field. Therefore, if we take $\psi_{\mathrm{C}}$ as a photon wave function, it should be interpreted as the energy-density amplitude of the photon field.

When $\boldsymbol{\theta}$ and $\boldsymbol{\varsigma}$ are the parameters of a rotation and a Lorentz boost, in terms of the rotations generators $\boldsymbol{J}$ and the boost generators $\boldsymbol{K}$, the Lorentz transformation of a spinor $\boldsymbol{F}$ can be expressed as $\boldsymbol{F}\rightarrow\exp(\mathrm{i}\boldsymbol{J}\cdot\boldsymbol{\theta}+\mathrm{i}\boldsymbol{K}\cdot\boldsymbol{\varsigma})\boldsymbol{F}$. In our case, one can show that $\boldsymbol{F}_{\mathrm{R}}=(\boldsymbol{E}+\mathrm{i}\boldsymbol{H})/2$ and $\boldsymbol{F}_{\mathrm{L}}=(\boldsymbol{E}-\mathrm{i}\boldsymbol{H})/2$ transforms as, respectively,

$$\boldsymbol{F}_{\mathrm{R}}\rightarrow\exp[\mathrm{i}\boldsymbol{\tau}\cdot(\boldsymbol{\theta}-\mathrm{i}\boldsymbol{\varsigma})]\boldsymbol{F}_{\mathrm{R}}\equiv D(\Lambda)\boldsymbol{F}_{\mathrm{R}}\text{, then }\ \boldsymbol{J}=\boldsymbol{\tau},\ \ \boldsymbol{K}=-\mathrm{i}\boldsymbol{\tau},\ \ (1,0)\,, \tag{14}$$

$$\boldsymbol{F}_{\mathrm{L}}\rightarrow\exp[\mathrm{i}\boldsymbol{\tau}\cdot(\boldsymbol{\theta}+\mathrm{i}\boldsymbol{\varsigma})]\boldsymbol{F}_{\mathrm{L}}\equiv \bar{D}(\Lambda)\boldsymbol{F}_{\mathrm{L}}\text{, then }\ \boldsymbol{J}=\boldsymbol{\tau},\ \ \boldsymbol{K}=\mathrm{i}\boldsymbol{\tau},\ \ (0,1)\,. \tag{15}$$

Here we have applied Eq. (4). Eqs. (14) and (15) imply that $D(\Lambda)=\exp[\mathrm{i}\boldsymbol{\tau}\cdot(\boldsymbol{\theta}-\mathrm{i}\boldsymbol{\varsigma})]$ and $\bar{D}(\Lambda)=\exp[\mathrm{i}\boldsymbol{\tau}\cdot(\boldsymbol{\theta}+\mathrm{i}\boldsymbol{\varsigma})]$ are the (1, 0) and (0, 1) representations of the group SL(2, C), respectively, they are inequivalent. Therefore, $\boldsymbol{F}_{\mathrm{R}}$ and $\boldsymbol{F}_{\mathrm{L}}$ are two different types of 3-component spinor, they transforms according to the (1, 0) and (0, 1) representations of the group SL(2, C), respectively, which implies that $\psi_{\mathrm{C}}$ is the $(1,0)\oplus(0,1)$ spinor (obviously, $\exp(\mathrm{i}\phi)\psi_{\mathrm{C}}$ is also the (1, 0) $\oplus$ (0, 1) spinor for $\phi\in[0,2\pi]$ and $\phi$ is a constant). In other words, under the Lorentz transformation of $x^{\mu}\rightarrow x'^{\mu}=\Lambda^{\mu}{}_{\nu}x^{\nu}$, the spinor $\psi_{\mathrm{C}}$ transforms as follows:

$$\psi_{\mathrm{C}}=\begin{pmatrix}\boldsymbol{F}_{\mathrm{R}}\\ \boldsymbol{F}_{\mathrm{L}}\end{pmatrix}\rightarrow\begin{pmatrix}D(\Lambda) & 0\\ 0 & \bar{D}(\Lambda)\end{pmatrix}\begin{pmatrix}\boldsymbol{F}_{\mathrm{R}}\\ \boldsymbol{F}_{\mathrm{L}}\end{pmatrix}=L_{\mathrm{C}}(\Lambda)\psi_{\mathrm{C}}, \tag{16}$$

where $L_{\mathrm{C}}(\Lambda)$ is the $(1,0)\oplus(0,1)$ representation of the group SL(2, C), i.e.,

$$L_{\mathrm{C}}(\Lambda)=\begin{pmatrix}D(\Lambda) & 0\\ 0 & \bar{D}(\Lambda)\end{pmatrix}=\begin{pmatrix}\exp[\mathrm{i}\boldsymbol{\tau}\cdot(\boldsymbol{\theta}-\mathrm{i}\boldsymbol{\varsigma})] & 0\\ 0 & \exp[\mathrm{i}\boldsymbol{\tau}\cdot(\boldsymbol{\theta}+\mathrm{i}\boldsymbol{\varsigma})]\end{pmatrix}. \tag{17}$$

As we know, the parity of the magnetic field $\boldsymbol{H}$ of an electric multipole of order ($l$, $m$) is $(-1)^{l}$, while the electric field $\boldsymbol{E}$ has parity $(-1)^{l+1}$, and the parities of the fields are just opposite to those of a magnetic multipole of the same order [62]. On the other hand, under space inversion (parity operation), the (1, 0) and (0, 1) representations become interchanged. For $\boldsymbol{F}_{\mathrm{R}}=(\boldsymbol{E}+\mathrm{i}\boldsymbol{H})/2$ and $\boldsymbol{F}_{\mathrm{L}}=(\boldsymbol{E}-\mathrm{i}\boldsymbol{H})/2$, under parity one has $\boldsymbol{F}_{\mathrm{R}}\leftrightarrow\pm\boldsymbol{F}_{\mathrm{L}}$ (note that for a constant $\phi\in[0,2\pi]$, $\exp(\mathrm{i}\phi)\boldsymbol{F}_{\mathrm{L}}$ is also the (0, 1) spinor), which implies that $\psi_{\mathrm{C}}$ transforms as follows (in Section 4 we will discuss parity operation further):

$$\psi_{\mathrm{C}}=\begin{pmatrix}\boldsymbol{F}_{\mathrm{R}}\\ \boldsymbol{F}_{\mathrm{L}}\end{pmatrix}\rightarrow\pm\begin{pmatrix}0 & I_{3\times3}\\ I_{3\times3} & 0\end{pmatrix}\begin{pmatrix}\boldsymbol{F}_{\mathrm{R}}\\ \boldsymbol{F}_{\mathrm{L}}\end{pmatrix}=\pm\beta_{\mathrm{C}}^{0}\psi_{\mathrm{C}}. \tag{18}$$

Therefore, once we consider parity, it is no longer sufficient to consider the 3-component spinors $\boldsymbol{F}_{\mathrm{R}}$ and $\boldsymbol{F}_{\mathrm{L}}$ separately, but the 6-component spinor $\psi_{\mathrm{C}}$. That is, the 6-component spinor $\psi_{\mathrm{C}}$ is an irreducible representation of the group SL(2, C) extended by parity. Eq. (9) presents a relation between the two spinors $\boldsymbol{F}_{\mathrm{R}}$ and $\boldsymbol{F}_{\mathrm{L}}$.

For photons with a 4D momentum $k^{\mu}=(\omega,\boldsymbol{k})$ satisfying $\omega=|\boldsymbol{k}|$, let $\hat{\boldsymbol{k}}=\boldsymbol{k}/\omega$, later we will show that

$$(\boldsymbol{\tau}\cdot\hat{\boldsymbol{k}})\boldsymbol{F}_{\mathrm{R}}=\boldsymbol{F}_{\mathrm{R}},\ \ (\boldsymbol{\tau}\cdot\hat{\boldsymbol{k}})\boldsymbol{F}_{\mathrm{L}}=-\boldsymbol{F}_{\mathrm{L}}. \tag{19}$$

Thus the spinors $\boldsymbol{F}_{\mathrm{R}}$ and $\boldsymbol{F}_{\mathrm{L}}$ are the eigenstates of helicity, the right-handed (left-handed) spinor having positive (negative) helicity.

The matrix $\beta_{\mathrm{C}}^{5}$ given by Eq. (12) is called a chirality operator (Eq. (12) provides its chiral representation). Define 6-component right- and left-handed spinors $\psi_{\mathrm{CR}}$, $\psi_{\mathrm{CL}}$:

$$\psi_{\mathrm{CR}}=\frac{1}{2}(1+\beta_{\mathrm{C}}^{5})\psi_{\mathrm{C}}=\begin{pmatrix}\boldsymbol{F}_{\mathrm{R}}\\ 0\end{pmatrix},\ \psi_{\mathrm{CL}}=\frac{1}{2}(1-\beta_{\mathrm{C}}^{5})\psi_{\mathrm{C}}=\begin{pmatrix}0\\ \boldsymbol{F}_{\mathrm{L}}\end{pmatrix}. \tag{20}$$

Obviously, they are also the solutions of Eq. (9). It is easy to show that

$$\beta_{\mathrm{C}}^{5}\psi_{\mathrm{CR}}=\psi_{\mathrm{CR}},\ \ \beta_{\mathrm{C}}^{5}\psi_{\mathrm{CL}}=-\psi_{\mathrm{CL}}. \tag{21}$$

Then the spinors $\psi_{\mathrm{CR}}$ and $\psi_{\mathrm{CL}}$ are the eigenstates of the chirality operator with the eigenvalues $\pm 1$ respectively.

Now let us introduce the standard representation $\psi_{\mathrm{S}}$ of the $(1,0)\oplus(0,1)$ spinor, it can be obtained from the chiral representation $\psi_{\mathrm{C}}$ via a unitary transformation:

$$\psi_{\mathrm{S}}=U\psi_{\mathrm{C}}=\frac{1}{\sqrt{2}}\begin{pmatrix}\boldsymbol{F}_{\mathrm{R}}+\boldsymbol{F}_{\mathrm{L}}\\ \boldsymbol{F}_{\mathrm{R}}-\boldsymbol{F}_{\mathrm{L}}\end{pmatrix}=\frac{1}{\sqrt{2}}\begin{pmatrix}\boldsymbol{E}\\ \mathrm{i}\boldsymbol{H}\end{pmatrix}. \tag{22}$$

where the unitary matrix $U$ satisfies

$$U=U^{\dagger}=U^{-1}=\frac{1}{\sqrt{2}}\begin{pmatrix}I_{3\times3} & I_{3\times3}\\ I_{3\times3} & -I_{3\times3}\end{pmatrix}. \tag{23}$$

In the standard representation, the matrices $M_{\mathrm{C}}$ $(=\beta_{\mathrm{C}}^{\mu},\ \boldsymbol{\alpha}_{\mathrm{C}},\ \boldsymbol{\Sigma}_{\mathrm{C}},\ \beta_{\mathrm{C}}^{5})$ given by Eqs. (10) and (12) are replaced with $M_{\mathrm{S}}=UM_{\mathrm{C}}U^{-1}$, where $M_{\mathrm{S}}=\beta_{\mathrm{S}}^{\mu},\ \boldsymbol{\alpha}_{\mathrm{S}},\ \boldsymbol{\Sigma}_{\mathrm{S}},\ \beta_{\mathrm{S}}^{5}$, one can obtain,

$$\beta_{\mathrm{S}}^{0}=\beta_{\mathrm{C}}^{5}=\begin{pmatrix}I_{3\times3} & 0\\ 0 & -I_{3\times3}\end{pmatrix},\ \ \boldsymbol{\beta}_{\mathrm{S}}=-\boldsymbol{\beta}_{\mathrm{C}}=\begin{pmatrix}0 & \boldsymbol{\tau}\\ -\boldsymbol{\tau} & 0\end{pmatrix}, \tag{24}$$

$$\boldsymbol{\alpha}_{\mathrm{S}}=\beta_{\mathrm{S}}^{0}\boldsymbol{\beta}_{\mathrm{S}}=\begin{pmatrix}0 & \boldsymbol{\tau}\\ \boldsymbol{\tau} & 0\end{pmatrix},\ \ \boldsymbol{\Sigma}_{\mathrm{S}}=\boldsymbol{\Sigma}_{\mathrm{C}}=\begin{pmatrix}\boldsymbol{\tau} & 0\\ 0 & \boldsymbol{\tau}\end{pmatrix},\ \ \beta_{\mathrm{S}}^{5}=\beta_{\mathrm{S5}}=\beta_{\mathrm{C}}^{0}=\begin{pmatrix}0 & I_{3\times3}\\ I_{3\times3} & 0\end{pmatrix}. \tag{25}$$

Seeing that $\boldsymbol{\Sigma}_{\mathrm{S}}=\boldsymbol{\Sigma}_{\mathrm{C}}$, the 3D spin matrix is denoted as $\boldsymbol{\Sigma}$. Likewise, under the Lorentz transformation of $x^\mu \to x'^\mu = \Lambda^\mu{}_\nu x^\nu$, the 6-component spinor $\psi_\mathrm{S}$ transforms according to $\psi_\mathrm{S} \to L_\mathrm{S}(\Lambda)\psi_\mathrm{S}$, where $L_\mathrm{S}(\Lambda)$ is equivalent to the (1, 0) ⊕ (0, 1) representation $L_\mathrm{C}(\Lambda)$, i.e.,

$$L_\mathrm{S} = UL_\mathrm{C}U^{-1} = \frac{1}{2}\begin{pmatrix} D(\Lambda)+\bar{D}(\Lambda) & D(\Lambda)-\bar{D}(\Lambda) \\ D(\Lambda)-\bar{D}(\Lambda) & D(\Lambda)+\bar{D}(\Lambda) \end{pmatrix}. \tag{26}$$

When $\psi(x)$ in Eq. (5) represents the photon field with $S^{\mu\nu}$ being the 4D spin tensor of the photon field, we will prove that, in the chiral representation one has $S_{lm} = \varepsilon_{lmn}\Sigma^n$ and $S^{0l} = \mathrm{i}\alpha_\mathrm{C}^l$, while in the standard representation one has $S_{lm} = \varepsilon_{lmn}\Sigma^n$ and $S^{0l} = \mathrm{i}\alpha_\mathrm{S}^l$, $l,m,n = 1,2,3$ (note that $\boldsymbol{\Sigma}_\mathrm{S} = \boldsymbol{\Sigma}_\mathrm{C} = \boldsymbol{\Sigma}$). Using $\omega^{l0} = -\omega^{0l} = \varsigma^l$ and $\omega^{lm} = -\varepsilon^{lmn}\theta_n$, Eq. (5) becomes

$$L = \exp(-\mathrm{i}\omega_{\mu\nu}S^{\mu\nu}/2) = \exp(\mathrm{i}\boldsymbol{\theta}\cdot\boldsymbol{\Sigma} + \boldsymbol{\varsigma}\cdot\boldsymbol{\alpha}). \tag{27}$$

To identify Eq. (27) with Eq. (17) or Eq. (26), let us prove that $L = L_\mathrm{C}$ for $S^{0l} = \mathrm{i}\alpha_\mathrm{C}^l$ (i.e., $\boldsymbol{\alpha} = \boldsymbol{\alpha}_\mathrm{C}$), or $L = L_\mathrm{S}$ for $S^{0l} = \mathrm{i}\alpha_\mathrm{S}^l$ (i.e., $\boldsymbol{\alpha} = \boldsymbol{\alpha}_\mathrm{S}$). In fact, using Eq. (12) one can show that

$$(\mathrm{i}\boldsymbol{\theta}\cdot\boldsymbol{\Sigma} + \boldsymbol{\varsigma}\cdot\boldsymbol{\alpha}_\mathrm{C})^l = \begin{pmatrix} (\mathrm{i}\boldsymbol{\theta}\cdot\boldsymbol{\tau} + \boldsymbol{\varsigma}\cdot\boldsymbol{\tau})^l & 0 \\ 0 & (\mathrm{i}\boldsymbol{\theta}\cdot\boldsymbol{\tau} - \boldsymbol{\varsigma}\cdot\boldsymbol{\tau})^l \end{pmatrix},\quad l = 1,2,3..., \tag{28}$$

which implies that

$$\exp(\mathrm{i}\boldsymbol{\theta}\cdot\boldsymbol{\Sigma} + \boldsymbol{\varsigma}\cdot\boldsymbol{\alpha}_\mathrm{C}) = \begin{pmatrix} \exp(\mathrm{i}\boldsymbol{\theta}\cdot\boldsymbol{\tau} + \boldsymbol{\varsigma}\cdot\boldsymbol{\tau}) & 0 \\ 0 & \exp(\mathrm{i}\boldsymbol{\theta}\cdot\boldsymbol{\tau} - \boldsymbol{\varsigma}\cdot\boldsymbol{\tau}) \end{pmatrix}. \tag{29}$$

Eqs. (17), (27) and (29) together imply that $L = \exp(-\mathrm{i}\omega_{\mu\nu}S^{\mu\nu}/2) = L_\mathrm{C}$ for $S^{0l} = \mathrm{i}\alpha_\mathrm{C}^l$, i.e.,

$$L_\mathrm{C} = \exp(\mathrm{i}\boldsymbol{\theta}\cdot\boldsymbol{\Sigma} + \boldsymbol{\varsigma}\cdot\boldsymbol{\alpha}_\mathrm{C}). \tag{30}$$

On the other hand, using $U\boldsymbol{A}_1\boldsymbol{A}_2...\boldsymbol{A}_mU^{-1} = U\boldsymbol{A}_1U^{-1}U\boldsymbol{A}_2U^{-1}U...U\boldsymbol{A}_mU^{-1}$, and seeing that $\boldsymbol{\Sigma}_\mathrm{S} = U\boldsymbol{\Sigma}_\mathrm{C}U^{-1} = \boldsymbol{\Sigma}_\mathrm{C} = \boldsymbol{\Sigma}$, $\boldsymbol{\alpha}_\mathrm{S} = U\boldsymbol{\alpha}_\mathrm{C}U^{-1}$, one has

$$L_\mathrm{S} = UL_\mathrm{C}U^{-1} = U\exp(\mathrm{i}\boldsymbol{\theta}\cdot\boldsymbol{\Sigma} + \boldsymbol{\varsigma}\cdot\boldsymbol{\alpha}_\mathrm{C})U^{-1} = \exp(\mathrm{i}\boldsymbol{\theta}\cdot\boldsymbol{\Sigma} + \boldsymbol{\varsigma}\cdot\boldsymbol{\alpha}_\mathrm{S}). \tag{31}$$

Therefore, $L = \exp(-\mathrm{i}\omega_{\mu\nu}S^{\mu\nu}/2) = L_\mathrm{S}$ for $S^{0l} = \mathrm{i}\alpha_\mathrm{S}^l$.

Let $\boldsymbol{a}=a\boldsymbol{n}=(a_1,a_2,a_3)$ with $a=\sqrt{a_1^2+a_2^2+a_3^2}$ and $\boldsymbol{n}\cdot\boldsymbol{n}=1$, and let $\boldsymbol{a}_{\mathrm{m}}$ (rather than $\boldsymbol{a}$ itself) denote the column-matrix form of the vector $\boldsymbol{a}$, that is

$$\boldsymbol{a}_{\mathrm{m}}=\begin{pmatrix}a_1\\a_2\\a_3\end{pmatrix},\quad \boldsymbol{a}_{\mathrm{m}}\boldsymbol{a}_{\mathrm{m}}^{\mathrm{T}}=\begin{pmatrix}a_1\\a_2\\a_3\end{pmatrix}\begin{pmatrix}a_1 & a_2 & a_3\end{pmatrix}. \tag{32}$$

It is easy to show that ( $l=1,2,...$ )

$$(\boldsymbol{a}\cdot\boldsymbol{\tau})^{2l+1}=a^{2l}(\boldsymbol{a}\cdot\boldsymbol{\tau}),\quad (\boldsymbol{a}\cdot\boldsymbol{\tau})^{2l+2}=a^{2l}(\boldsymbol{a}\cdot\boldsymbol{\tau})^2,\quad (\boldsymbol{a}\cdot\boldsymbol{\tau})^2=a^2-\boldsymbol{a}_{\mathrm{m}}\boldsymbol{a}_{\mathrm{m}}^{\mathrm{T}}. \tag{33}$$

It follows from Eq. (33) that

$$\exp(\pm\mathrm{i}\boldsymbol{a}\cdot\boldsymbol{\tau})=\cos a\pm\mathrm{i}\frac{(\boldsymbol{a}\cdot\boldsymbol{\tau})}{a}\sin a+\frac{\boldsymbol{a}_{\mathrm{m}}\boldsymbol{a}_{\mathrm{m}}^{\mathrm{T}}}{a^2}(1-\cos a). \tag{34}$$

Note that as $a\to 0$, one has $\boldsymbol{a}_{\mathrm{m}}\boldsymbol{a}_{\mathrm{m}}^{\mathrm{T}}\sim a^2\to 0$ and $(1-\cos a)/a^2\to 1/2$, such that $(1-\cos a)\boldsymbol{a}_{\mathrm{m}}\boldsymbol{a}_{\mathrm{m}}^{\mathrm{T}}/a^2\sim a^2\to 0$. For our purpose, let $\boldsymbol{a}=\boldsymbol{\theta}-\mathrm{i}\boldsymbol{\varsigma}$ (its complex conjugate is $\boldsymbol{a}^*=\boldsymbol{\theta}+\mathrm{i}\boldsymbol{\varsigma}$) with $a_l=\theta_l-\mathrm{i}\varsigma_l$ ( $l=1,2,3$ ). For simplicity, let $\boldsymbol{a}=(a_1,0,0)$, $a=a_1=\theta_1-\mathrm{i}\varsigma_1$, and then $\exp(\mathrm{i}\boldsymbol{a}\cdot\boldsymbol{\tau})=\exp(\mathrm{i}a_1\tau_1)$, using Eqs. (11) , (17) and (34), one has

$$\exp(\mathrm{i}a_1\tau_1)=\begin{pmatrix}1 & 0 & 0\\ 0 & \cos a_1 & \sin a_1\\ 0 & -\sin a_1 & \cos a_1\end{pmatrix},\quad L_{\mathrm{C}}=\begin{pmatrix}\exp(\mathrm{i}a_1\tau_1) & 0\\ 0 & \exp(\mathrm{i}a_1^*\tau_1)\end{pmatrix}. \tag{35}$$

Let us discuss Eq. (35) as follows:

1) For a rotation about the $x$ axis through the angle $\theta_1$, one has $\boldsymbol{\theta}=(\theta_1,0,0)$, $\boldsymbol{\varsigma}=(0,0,0)$, for the moment $a_1=a_1^*=\theta_1$, and Eq. (35) becomes

$$\exp(\mathrm{i}\theta_1\tau_1)=\begin{pmatrix}1 & 0 & 0\\ 0 & \cos\theta_1 & \sin\theta_1\\ 0 & -\sin\theta_1 & \cos\theta_1\end{pmatrix},\quad L_{\mathrm{C}}=\begin{pmatrix}\exp(\mathrm{i}\theta_1\tau_1) & 0\\ 0 & \exp(\mathrm{i}\theta_1\tau_1)\end{pmatrix}. \tag{36}$$

In fact, under this rotation transformation, starting from the transformation rules of the electromagnetic fields $\boldsymbol{E}$ and $\boldsymbol{H}$, one can also show that the chiral representation $\psi_{\mathrm{C}}$ given by Eq. (7) transforms in the manner of $\psi_{\mathrm{C}} \to L_{\mathrm{C}}\psi_{\mathrm{C}}$, which accords with Eq. (16).

2) In a pure Lorentz transformation with the relative velocity $v_1$ along the $x$ axis satisfying $\tanh\varsigma_1 = v_1$, one has $\boldsymbol{\theta} = (0,0,0)$, $\varsigma = (\varsigma_1, 0, 0)$, and $a_1 = -a_1^* = -\mathrm{i}\varsigma_1$, using $\cos(\pm\mathrm{i}\varsigma_1) = \cosh\varsigma_1$ and $\sin(\pm\mathrm{i}\varsigma_1) = \pm\mathrm{i}\sinh\varsigma_1$, one can obtain from Eq. (35):

$$\exp(\pm\varsigma_1\tau_1) = \begin{pmatrix} 1 & 0 & 0 \\ 0 & \cosh\varsigma_1 & \mp\mathrm{i}\sinh\varsigma_1 \\ 0 & \pm\mathrm{i}\sinh\varsigma_1 & \cosh\varsigma_1 \end{pmatrix}, \quad L_{\mathrm{C}} = \begin{pmatrix} \exp(\varsigma_1\tau_1) & 0 \\ 0 & \exp(-\varsigma_1\tau_1) \end{pmatrix}. \tag{37}$$

In fact, under such a pure Lorentz transformation, starting from the transformation rules of the electromagnetic fields $\boldsymbol{E}$ and $\boldsymbol{H}$, one can also show that the chiral representation $\psi_{\mathrm{C}}$ given by Eq. (7) transforms in the manner of $\psi_{\mathrm{C}} \to L_{\mathrm{C}}\psi_{\mathrm{C}}$, which accords with Eq. (16).

To pave the way for our later investigations, let us present several useful formulae. Let $a^\mu = (a^0, \boldsymbol{a})$ and $b^\mu = (b^0, \boldsymbol{b})$ be two 4D vectors, the column matrix forms of $\boldsymbol{a}$ and $\boldsymbol{b}$ are denoted as $\boldsymbol{a}_{\mathrm{m}} = \begin{pmatrix} a_1 & a_2 & a_3 \end{pmatrix}^{\mathrm{T}}$, $\boldsymbol{b}_{\mathrm{m}} = \begin{pmatrix} b_1 & b_2 & b_3 \end{pmatrix}^{\mathrm{T}}$, respectively, one can prove that, for $\beta^\mu = \beta^\mu_{\mathrm{C}}$, $\beta^\mu_{\mathrm{S}}$, $\boldsymbol{\alpha} = \boldsymbol{\alpha}_{\mathrm{C}}$, $\boldsymbol{\alpha}_{\mathrm{S}}$, and $\boldsymbol{\Sigma} = \boldsymbol{\Sigma}_{\mathrm{C}} = \boldsymbol{\Sigma}_{\mathrm{S}}$, one has

$$(\boldsymbol{\tau}\cdot\boldsymbol{a})(\boldsymbol{\tau}\cdot\boldsymbol{b}) = \boldsymbol{a}\cdot\boldsymbol{b} + \mathrm{i}\boldsymbol{\tau}\cdot(\boldsymbol{a}\times\boldsymbol{b}) - \boldsymbol{a}_{\mathrm{m}}\boldsymbol{b}_{\mathrm{m}}^{\mathrm{T}}, \quad \boldsymbol{a}_{\mathrm{m}}^{\mathrm{T}}\boldsymbol{\tau}\boldsymbol{b}_{\mathrm{m}} = -\mathrm{i}(\boldsymbol{a}\times\boldsymbol{b}), \tag{38}$$

$$(\beta^\mu a_\mu)(\beta^\nu b_\nu) = -a^\mu b_\mu - \mathrm{i}\boldsymbol{\Sigma}\cdot(\boldsymbol{a}\times\boldsymbol{b}) + \boldsymbol{\alpha}\cdot(\boldsymbol{a}b^0 - a^0\boldsymbol{b}) + I_{2\times2}\otimes\boldsymbol{a}_{\mathrm{m}}\boldsymbol{b}_{\mathrm{m}}^{\mathrm{T}}. \tag{39}$$

Here we have left out the unit matrices $I_{3\times3}$ and $I_{6\times6}$ that are equivalent to 1 (the same below).

All the known photon wave functions cannot possess all the properties of the Schrödinger wave functions of nonrelativistic wave mechanics, such that photon wave function has ever been a controversial concept. However, this is a common property when quantum mechanics is extended from nonrelativistic case to relativistic one (the exact theory of the latter is quantum field theory). That is, in quantum field theory, the concept of wave function should be replaced with that of field operator. Nevertheless, the quantum average of a field operator can still be regarded as a wave function, but its physical meanings have been generalized: it may be a probability (-density) amplitude, charge-density amplitude, or energy-density amplitude, etc. In our opinion, it is unessential whether the $(1,0)\oplus(0,1)$ spinor can be regarded as a photon wave function or not, it is just a matter of appellation. It is a field operator from the point of view of quantum field theory.

**3. The quantization of the photon field based on the $(1,0)\oplus(0,1)$ spinor description**

When the photon field is described by a 4D electromagnetic potential that transforms according to the (1/2, 1/2) representation of the group SL(2, C), it cannot be the eigenstates of helicity $\pm 1$ (i.e., it is impossible to construct a four-vector field for massless particles of helicity $\pm 1$ [3]). On the other hand, when the photon field is described by the $6\times 1$ spinor $\psi=\psi_{\mathrm{C}},\ \psi_{\mathrm{S}}$ that transforms according to the $(1,0)\oplus(0,1)$ representation of the group SL(2, C), it is no problem in constructing a field for a particle of mass zero and helicity $\pm 1$. In other words, the $(1,0)\oplus(0,1)$ spinor field exhibits simpler Lorentz transformation properties than a four-vector field. For convenience, let the photon field below $\psi=\psi_{\mathrm{S}},\ \psi_{\mathrm{C}}$, and the related matrices $M=M_{\mathrm{S}},\ M_{\mathrm{C}}$, where

a) In the standard representation: $\psi=\psi_{\mathrm{S}}$, $M_{\mathrm{S}}=\beta_{\mathrm{S}}^{\mu},\ \boldsymbol{\alpha}_{\mathrm{S}},\ \boldsymbol{\Sigma}_{\mathrm{S}},\ \beta_{\mathrm{S}}^{5}$, and

$$\psi_{\mathrm{S}}=\frac{1}{\sqrt{2}}\begin{pmatrix}\boldsymbol{E}\\ \mathrm{i}\boldsymbol{H}\end{pmatrix},\ \ \beta_{\mathrm{S}}^{0}=\begin{pmatrix}I_{3\times3} & 0\\ 0 & -I_{3\times3}\end{pmatrix},\ \ \boldsymbol{\beta}_{\mathrm{S}}=\begin{pmatrix}0 & \boldsymbol{\tau}\\ -\boldsymbol{\tau} & 0\end{pmatrix}, \tag{40-1}$$

$$\boldsymbol{\alpha}_{\mathrm{S}}=\begin{pmatrix}0 & \boldsymbol{\tau}\\ \boldsymbol{\tau} & 0\end{pmatrix},\ \ \boldsymbol{\Sigma}_{\mathrm{S}}=\boldsymbol{\Sigma}=\begin{pmatrix}\boldsymbol{\tau} & 0\\ 0 & \boldsymbol{\tau}\end{pmatrix},\ \ \beta_{\mathrm{S}}^{5}=\begin{pmatrix}0 & I_{3\times3}\\ I_{3\times3} & 0\end{pmatrix}. \tag{40-2}$$

b) In the chiral representation: $\psi=\psi_{\mathrm{C}}=U\psi_{\mathrm{S}}$, $M_{\mathrm{C}}=UM_{\mathrm{S}}U^{-1}=\beta_{\mathrm{C}}^{\mu},\ \boldsymbol{\alpha}_{\mathrm{C}},\ \boldsymbol{\Sigma}_{\mathrm{C}},\ \beta_{\mathrm{C}}^{5}$, and

$$\psi_{\mathrm{C}}=\begin{pmatrix}\boldsymbol{F}_{\mathrm{R}}\\ \boldsymbol{F}_{\mathrm{L}}\end{pmatrix}=\frac{1}{2}\begin{pmatrix}\boldsymbol{E}+\mathrm{i}\boldsymbol{H}\\ \boldsymbol{E}-\mathrm{i}\boldsymbol{H}\end{pmatrix},\ \ \beta_{\mathrm{C}}^{0}=\begin{pmatrix}0 & I_{3\times3}\\ I_{3\times3} & 0\end{pmatrix},\ \ \boldsymbol{\beta}_{\mathrm{C}}=\begin{pmatrix}0 & -\boldsymbol{\tau}\\ \boldsymbol{\tau} & 0\end{pmatrix}, \tag{41-1}$$

$$\boldsymbol{\alpha}_{\mathrm{C}}=\begin{pmatrix}\boldsymbol{\tau} & 0\\ 0 & -\boldsymbol{\tau}\end{pmatrix},\ \boldsymbol{\Sigma}_{\mathrm{C}}=\boldsymbol{\Sigma}=\begin{pmatrix}\boldsymbol{\tau} & 0\\ 0 & \boldsymbol{\tau}\end{pmatrix},\ \beta_{\mathrm{C}}^{5}=\begin{pmatrix}I_{3\times3} & 0\\ 0 & -I_{3\times3}\end{pmatrix}, \tag{41-2}$$

where the unitary matrix $U$ is given by Eq. (23). For the moment the Dirac-equation is written in the unified form ( $\hat{H}=-\mathrm{i}\boldsymbol{\alpha}\cdot\nabla$ ):

$$\mathrm{i}\beta^{\mu}\partial_{\mu}\psi(x)=0\text{, or }\ \mathrm{i}\partial_{t}\psi(x)=\hat{H}\psi(x). \tag{42}$$

Using the transverse conditions given by Eq. (6-2) one can obtain $\partial^{\mu}\partial_{\mu}\psi(x)=0$ from Eq. (42). Substituting $\varphi(k)\exp(-\mathrm{i}k\cdot x)$ into Eq. (42) one has $\det(\omega-\boldsymbol{\alpha}\cdot\boldsymbol{k})=0$, i.e.,

$$\det(\omega-\boldsymbol{\alpha}\cdot\boldsymbol{k})=\omega^{2}[\omega^{2}-(k_{1}^{2}+k_{2}^{2}+k_{3}^{2})]^{2}=0. \tag{43}$$

It follows from Eq. (43) that $\omega^2$ has three solutions, and then let us denote $\omega^2$ as $\omega_{\lambda}^{2}=\omega_{\lambda}^{2}(\boldsymbol{k})$ with $\lambda=0,\pm1$. Form Eq. (43) we obtain the dispersion relations as follows:

$$\omega_{\lambda}^{2}=\boldsymbol{k}^{2}=k_{1}^{2}+k_{2}^{2}+k_{3}^{2}\ \text{ for }\ \lambda=\pm1;\ \ \omega_{\lambda}^{2}=0\ \text{ for }\ \lambda=0. \tag{44}$$

As shown later, the $\lambda=\pm1$ solutions describe the transverse photons, while the $\lambda=0$ solution describes the admixture of the longitudinal and scalar photons, and the latter would be discarded once the transverse conditions are taken into account.

Before seeking for the general solution of Eq. (42), let us first study electromagnetic field intensities $\boldsymbol{E}$ and $\boldsymbol{H}$ starting from a 4D electromagnetic potential. From now on, field quantities are written in the form of a field operator. In the Lorentz gauge condition, a 4D electromagnetic potential $\hat{A}^{\mu}(x)=(\hat{A}^{0},\hat{\boldsymbol{A}})$ can be expanded as:

$$\hat{A}^{\mu}(x)=\int\frac{\mathrm{d}^{3}k}{\sqrt{2|\boldsymbol{k}|(2\pi)^{3}}}\sum_{s=0}^{3}\eta^{\mu}(\boldsymbol{k},s)\{\hat{c}(\boldsymbol{k},s)\exp[-\mathrm{i}(\omega_{s}t-\boldsymbol{k}\cdot\boldsymbol{x})]+\hat{c}^{\dagger}(\boldsymbol{k},s)\exp[\mathrm{i}(\omega_{s}t-\boldsymbol{k}\cdot\boldsymbol{x})]\}, \tag{45}$$

where $\eta^{\mu}(\boldsymbol{k},s)$ ( $s=0,1,2,3$ , $\mu=0,1,2,3$ ) are four 4D polarization vectors, the four indices of $s$=0,1,2,3 describe four kinds of photons, respectively. In contrast to the traditional expression, here the frequencies of the four kinds of photons are temporarily denoted as $\omega_s = \omega_s(\boldsymbol{k})$ for $s = 0,1,2,3$, respectively. One can choose

$$\begin{cases} \eta^{\mu}(\boldsymbol{k},0) = (1,0,0,0), & \eta^{\mu}(\boldsymbol{k},1) = (0,\boldsymbol{\eta}(\boldsymbol{k},1)) \\ \eta^{\mu}(\boldsymbol{k},2) = (0,\boldsymbol{\eta}(\boldsymbol{k},2)), & \eta^{\mu}(\boldsymbol{k},3) = (0,\boldsymbol{\eta}(\boldsymbol{k},3)) \end{cases}, \tag{46}$$

where $\boldsymbol{\eta}(\boldsymbol{k},i)$ ($i = 1,2,3$) are three 3D linear polarization vectors, they satisfy the relation of $\boldsymbol{\eta}(\boldsymbol{k},1) \times \boldsymbol{\eta}(\boldsymbol{k},2) = \boldsymbol{\eta}(\boldsymbol{k},3) = \boldsymbol{k}/|\boldsymbol{k}|$. Obviously, $\boldsymbol{\eta}(\boldsymbol{k},1)$ and $\boldsymbol{\eta}(\boldsymbol{k},2)$ (perpendicular to $\boldsymbol{k}$) are two transverse polarization vectors, while $\boldsymbol{\eta}(\boldsymbol{k},3)$ (parallel to $\boldsymbol{k}$) is the longitudinal polarization vector. The column-matrix forms of $\boldsymbol{\eta}(\boldsymbol{k},i)$ are denoted as $\boldsymbol{\varepsilon}(\boldsymbol{k},i)$ for $i = 1,2,3$, respectively, which can be expressed as,

$$\boldsymbol{\varepsilon}(\boldsymbol{k},1) = \frac{1}{|\boldsymbol{k}|}\begin{pmatrix} \dfrac{k_1^2 k_3 + k_2^2|\boldsymbol{k}|}{k_1^2 + k_2^2} \\ \dfrac{k_1 k_2 (k_3 - |\boldsymbol{k}|)}{k_1^2 + k_2^2} \\ -k_1 \end{pmatrix}, \quad \boldsymbol{\varepsilon}(\boldsymbol{k},2) = \frac{1}{|\boldsymbol{k}|}\begin{pmatrix} \dfrac{k_1 k_2 (k_3 - |\boldsymbol{k}|)}{k_1^2 + k_2^2} \\ \dfrac{k_1^2|\boldsymbol{k}| + k_2^2 k_3}{k_1^2 + k_2^2} \\ -k_2 \end{pmatrix}, \quad \boldsymbol{\varepsilon}(\boldsymbol{k},3) = \frac{1}{|\boldsymbol{k}|}\begin{pmatrix} k_1 \\ k_2 \\ k_3 \end{pmatrix}. \tag{47}$$

It follows from $\boldsymbol{a}_{\mathrm{m}}^{\mathrm{T}} \boldsymbol{\tau} \boldsymbol{b}_{\mathrm{m}} = -\mathrm{i}(\boldsymbol{a} \times \boldsymbol{b})$ given by Eq. (38) that

$$\boldsymbol{\varepsilon}^{\dagger}(\boldsymbol{k},i)\boldsymbol{\tau}\boldsymbol{\varepsilon}(\boldsymbol{k},j) = -\mathrm{i}[\boldsymbol{\eta}(\boldsymbol{k},i) \times \boldsymbol{\eta}(\boldsymbol{k},j)], \quad i,j = 1,2,3. \tag{48}$$

Using Eqs. (11), (47) and (48) one can prove that:

$$\boldsymbol{\varepsilon}^{\dagger}(\boldsymbol{k},i)\boldsymbol{\varepsilon}(\boldsymbol{k},j) = \delta_{ij}, \quad \sum\nolimits_{i=1}^{3} \boldsymbol{\varepsilon}(\boldsymbol{k},i)\boldsymbol{\varepsilon}^{\dagger}(\boldsymbol{k},i) = I_{3\times 3}, \quad i,j = 1,2,3. \tag{49-1}$$

$$\boldsymbol{\varepsilon}^{\dagger}(\boldsymbol{k},1)\boldsymbol{\varepsilon}(-\boldsymbol{k},1) = -\boldsymbol{\varepsilon}^{\dagger}(\boldsymbol{k},2)\boldsymbol{\varepsilon}(-\boldsymbol{k},2) = (k_2^2 - k_1^2)/(k_1^2 + k_2^2), \tag{49-2}$$

$$\boldsymbol{\varepsilon}^{\dagger}(\boldsymbol{k},1)\boldsymbol{\varepsilon}(-\boldsymbol{k},2) = \boldsymbol{\varepsilon}^{\dagger}(\boldsymbol{k},2)\boldsymbol{\varepsilon}(-\boldsymbol{k},1) = -2k_1 k_2/(k_1^2 + k_2^2), \tag{49-3}$$

$$\boldsymbol{\varepsilon}^{\dagger}(\boldsymbol{k},1)\boldsymbol{\tau}\boldsymbol{\varepsilon}(\boldsymbol{k},2) = -\boldsymbol{\varepsilon}^{\dagger}(\boldsymbol{k},2)\boldsymbol{\tau}\boldsymbol{\varepsilon}(\boldsymbol{k},1) = -\mathrm{i}\,\boldsymbol{k}/\omega, \tag{49-4}$$

$$\boldsymbol{\varepsilon}^{\dagger}(\boldsymbol{k},1)\boldsymbol{\tau}\boldsymbol{\varepsilon}(\boldsymbol{k},1) = \boldsymbol{\varepsilon}^{\dagger}(\boldsymbol{k},2)\boldsymbol{\tau}\boldsymbol{\varepsilon}(\boldsymbol{k},2) = (0,0,0), \tag{49-5}$$

$$\boldsymbol{\varepsilon}^{\dagger}(\boldsymbol{k},1)\boldsymbol{\tau}\boldsymbol{\varepsilon}(-\boldsymbol{k},2) = \boldsymbol{\varepsilon}^{\dagger}(\boldsymbol{k},2)\boldsymbol{\tau}\boldsymbol{\varepsilon}(-\boldsymbol{k},1) = \mathrm{i}\boldsymbol{k}(k_2^2 - k_1^2)/\omega(k_1^2 + k_2^2), \tag{49-6}$$

$$\boldsymbol{\varepsilon}^{\dagger}(\boldsymbol{k},1)\boldsymbol{\tau}\boldsymbol{\varepsilon}(-\boldsymbol{k},1) = -\boldsymbol{\varepsilon}^{\dagger}(\boldsymbol{k},2)\boldsymbol{\tau}\boldsymbol{\varepsilon}(-\boldsymbol{k},2) = 2\mathrm{i}\boldsymbol{k}k_1 k_2/\omega(k_1^2 + k_2^2). \tag{49-7}$$

It is important to notice that, if $\boldsymbol{\eta}(\boldsymbol{k},i)$ ($i=1,2,3$) were three 3D circular polarization vectors, $A^{\mu}(x)$ would no longer be a 4D vector [3], which is related to the fact that the (1/2, 1/2) representation of the group SL(2, C) cannot be the eigenstates of helicity $\pm 1$. Even so, in the present case, $F_{\mu\nu}=\partial_{\mu}A_{\nu}-\partial_{\nu}A_{\mu}$ is still an antisymmetric tensor.

Substituting Eq. (45) into $\hat{\boldsymbol{E}}=-\nabla\hat{A}^{0}-\partial\hat{\mathbf{A}}/\partial t$ and $\hat{\boldsymbol{H}}=\nabla\times\hat{\mathbf{A}}$ one can prove that (note that $\boldsymbol{\eta}(\boldsymbol{k},3)=\boldsymbol{k}/|\boldsymbol{k}|$, and $\boldsymbol{k}\times\boldsymbol{\eta}(\boldsymbol{k},3)=0$)

$$\hat{\boldsymbol{E}}=\hat{\boldsymbol{E}}_{\perp}+\hat{\boldsymbol{E}}_{\|},\quad \hat{\boldsymbol{H}}=\hat{\boldsymbol{H}}_{\perp}, \tag{50}$$

where

$$\hat{\boldsymbol{E}}_{\perp}=\int\frac{\mathrm{d}^{3}k}{\sqrt{2|\boldsymbol{k}|(2\pi)^{3}}}\sum_{i=1}^{2}\omega_{i}\boldsymbol{\eta}(\boldsymbol{k},i)\{\mathrm{i}\hat{c}(\boldsymbol{k},i)\exp[-\mathrm{i}(\omega_{i}t-\boldsymbol{k}\cdot\boldsymbol{x})]-\mathrm{i}\hat{c}^{\dagger}(\boldsymbol{k},i)\exp[\mathrm{i}(\omega_{i}t-\boldsymbol{k}\cdot\boldsymbol{x})]\},$$

$$\begin{aligned}\hat{\boldsymbol{E}}_{\|}=&\int\frac{\mathrm{d}^{3}k}{\sqrt{2|\boldsymbol{k}|(2\pi)^{3}}}\{[|\boldsymbol{k}|\boldsymbol{\eta}(\boldsymbol{k},3)][-\mathrm{i}\hat{c}(\boldsymbol{k},0)\exp[-\mathrm{i}(\omega_{0}t-\boldsymbol{k}\cdot\boldsymbol{x})]+\mathrm{i}\hat{c}^{\dagger}(\boldsymbol{k},0)\exp[\mathrm{i}(\omega_{0}t-\boldsymbol{k}\cdot\boldsymbol{x})]\}\\&+\int\frac{\mathrm{d}^{3}k}{\sqrt{2|\boldsymbol{k}|(2\pi)^{3}}}[\omega_{3}\boldsymbol{\eta}(\boldsymbol{k},3)]\{\mathrm{i}\hat{c}(\boldsymbol{k},3)\exp[-\mathrm{i}(\omega_{3}t-\boldsymbol{k}\cdot\boldsymbol{x})]-\mathrm{i}\hat{c}^{\dagger}(\boldsymbol{k},3)\exp[\mathrm{i}(\omega_{3}t-\boldsymbol{k}\cdot\boldsymbol{x})]\}\end{aligned},$$

$$\hat{\boldsymbol{H}}_{\perp}=\int\frac{\mathrm{d}^{3}k}{\sqrt{2|\boldsymbol{k}|(2\pi)^{3}}}\sum_{i=1}^{2}[\boldsymbol{k}\times\boldsymbol{\eta}(\boldsymbol{k},i)]\{\mathrm{i}\hat{c}(\boldsymbol{k},i)\exp[-\mathrm{i}(\omega_{i}t-\boldsymbol{k}\cdot\boldsymbol{x})]-\mathrm{i}\hat{c}^{\dagger}(\boldsymbol{k},i)\exp[\mathrm{i}(\omega_{i}t-\boldsymbol{k}\cdot\boldsymbol{x})]\}. \tag{51}$$

Obviously, the transverse fields $\hat{\boldsymbol{E}}_{\perp}$ and $\hat{\boldsymbol{H}}_{\perp}$, related to the polarization vectors $\boldsymbol{\eta}(\boldsymbol{k},i)$ ($i=1,\ 2$) and satisfying the transversality conditions Eq. (6-2), are radiation fields, they consist of two kinds of transverse photons (i.e., the $s=1,\ 2$ photons); while the longitudinal field $\hat{\boldsymbol{E}}_{\|}$, related to the polarization vectors $\boldsymbol{\eta}(\boldsymbol{k},3)$ and failing to satisfy the transversality condition, represents a non-radiation field, it consists of the longitudinal ($s$=3) and scalar ($s$=0) photons.

The frequency of the radiation fields depends on the one of a radiation source, it is advisable to assume that $\omega_s = \omega$ for $s = 1, 2$. On the other hand, a non-radiation field stems from an electrostatic field carried by charged particles, where the electrostatic field may move following the charged particles. Once the charged particles keep an uniform motion, the frequencies of $\hat{\boldsymbol{E}}_{\parallel}$ vanish, that is, $\omega_s = 0$ for $s = 0, 3$, which in agreement with Eq. (44). However, once we consider the transverse conditions, the longitudinal field $\hat{\boldsymbol{E}}_{\parallel}$ should be discarded.

Now let us come back to the traditional formalism in which $\omega_s = \omega = |\boldsymbol{k}|$ for $s = 0,1,2,3$, Eq. (45) becomes ( $k \cdot x = k^{\mu} x_{\mu} = -\omega t + \boldsymbol{k} \cdot \boldsymbol{x}$ )

$$\hat{A}^{\mu}(x) = \int \frac{\mathrm{d}^3 k}{\sqrt{2\omega(2\pi)^3}} \sum_{s=0}^{3} \eta^{\mu}(\boldsymbol{k}, s)[\hat{c}(\boldsymbol{k}, s)\exp(\mathrm{i}k \cdot x) + \hat{c}^{\dagger}(\boldsymbol{k}, s)\exp(-\mathrm{i}k \cdot x)], \quad (52)$$

and Eq. (51) becomes (using $\boldsymbol{\eta}(\boldsymbol{k}, 3) = \boldsymbol{k}/|\boldsymbol{k}| = \boldsymbol{k}/\omega$ )

$$\hat{\boldsymbol{E}}_{\perp} = \int \frac{\sqrt{\omega}\mathrm{d}^3 k}{\sqrt{2(2\pi)^3}} \sum_{i=1}^{2} \{\boldsymbol{\eta}(\boldsymbol{k}, i)[\hat{b}(\boldsymbol{k}, i)\exp(\mathrm{i}k \cdot x) + \hat{b}^{\dagger}(\boldsymbol{k}, i)\exp(-\mathrm{i}k \cdot x)]\}, \quad (53)$$

$$\hat{\boldsymbol{E}}_{\parallel} = \int \frac{\sqrt{\omega}\mathrm{d}^3 k}{\sqrt{2(2\pi)^3}} \{\boldsymbol{\eta}(\boldsymbol{k}, 3)[\hat{b}(\boldsymbol{k}, 3)\exp(\mathrm{i}k \cdot x) + \hat{b}^{\dagger}(\boldsymbol{k}, 3)\exp(-\mathrm{i}k \cdot x)]\}, \quad (54)$$

$$\hat{\boldsymbol{H}}_{\perp} = \int \frac{\sqrt{\omega}\mathrm{d}^3 k}{\sqrt{2(2\pi)^3}} \sum_{i=1}^{2} \{[\boldsymbol{\eta}(\boldsymbol{k}, 3) \times \boldsymbol{\eta}(\boldsymbol{k}, i)][\hat{b}(\boldsymbol{k}, i)\exp(\mathrm{i}k \cdot x) + \hat{b}^{\dagger}(\boldsymbol{k}, i)\exp(-\mathrm{i}k \cdot x)]\}, \quad (55)$$

where the relations between the operators $\hat{b}(\boldsymbol{k}, i)$ and $\hat{c}(\boldsymbol{k}, s)$ are,

$$\hat{b}(\boldsymbol{k}, 1) = \mathrm{i}\hat{c}(\boldsymbol{k}, 1), \quad \hat{b}(\boldsymbol{k}, 2) = \mathrm{i}\hat{c}(\boldsymbol{k}, 2), \quad \hat{b}(\boldsymbol{k}, 3) = -\mathrm{i}[\hat{c}(\boldsymbol{k}, 0) - \hat{c}(\boldsymbol{k}, 3)]. \quad (56)$$

Therefore, when the electromagnetic field is described by the 4D electromagnetic potential $\hat{A}^{\mu}(x)$, there involves four 4D polarization vectors $\eta^{\mu}(\boldsymbol{k}, s)$ ( $s = 0,1,2,3$ ) together describing four kinds of photons; while described by the electromagnetic field intensities $\hat{\boldsymbol{E}}$ and $\hat{\boldsymbol{H}}$, there involves three 3D polarization vectors $\boldsymbol{\eta}(\boldsymbol{k}, i)$ ( $i = 1,2,3$ ). Moreover, Eq.

(56) shows that the $i=1,\ 2$ solutions of $\hat{b}(\boldsymbol{k},i)$ describe two kinds of transverse photons ($\hat{c}(\boldsymbol{k},s)$ for $s=1,\ 2$), while the $i=3$ solution corresponds to the admixture of the longitudinal ($s$=3) and scalar ($s$=0) photons.

Eq. (6) describes the Maxwell equations in Minkowski vacuum. Here, however, let us first consider that Maxwell equations with the current density of $j_e^\mu=(\rho_e,\boldsymbol{j}_e)$, and then

$$\rho_e=\nabla\cdot\hat{\boldsymbol{E}}\,,\quad \boldsymbol{j}_e=\nabla\times\hat{\boldsymbol{H}}-\partial\hat{\boldsymbol{E}}/\partial t\,. \tag{57}$$

Substituting Eqs. (50) and (53)-(55) into Eq. (57), one has

$$\rho_e=\int\frac{\omega\sqrt{\omega}\mathrm{d}^3k}{\sqrt{2(2\pi)^3}}[\mathrm{i}\hat{b}(\boldsymbol{k},3)\exp(\mathrm{i}k\cdot x)-\mathrm{i}\hat{b}^\dagger(\boldsymbol{k},3)\exp(-\mathrm{i}k\cdot x)]\,, \tag{58}$$

$$\boldsymbol{j}_e=\int\frac{\omega\sqrt{\omega}\mathrm{d}^3k}{\sqrt{2(2\pi)^3}}\{\boldsymbol{\eta}(\boldsymbol{k},3)[\mathrm{i}\hat{b}(\boldsymbol{k},3)\exp(\mathrm{i}k\cdot x)-\mathrm{i}\hat{b}^\dagger(\boldsymbol{k},3)\exp(-\mathrm{i}k\cdot x)]\}\,. \tag{59}$$

Using Eqs. (58) and (59) one can obtain $\partial\rho_e/\partial t+\nabla\cdot\boldsymbol{j}_e=0$. Eqs. (56), (58) and (59) together imply that the longitudinal and scalar photons are related to the current density.

Now let us come back to the case of $j_e^\mu=0$. For the moment, the transversality condition of $\nabla\cdot\hat{\boldsymbol{E}}=0$ requires us to throw away $\hat{\boldsymbol{E}}_{\|}$ given by Eq. (54), such that we only take into account the transverse photons by taking $\hat{\boldsymbol{E}}=\hat{\boldsymbol{E}}_{\perp}$. Therefore, in terms of the $(1,0)\oplus(0,1)$ spinor, one can develop the theory of the quantized photon field without resorting to the Gupta-Bleuler method. According to the Gupta-Bleuler method, in terms of the 4D electromagnetic potential $\hat{A}^\mu(x)$ given by Eq. (52), the theory of the quantized electromagnetic field is described in terms of a Hilbert space with an indefinite metric and with use of the assumption that, only those state vectors (say, $|\Phi\rangle$) are admitted for which the expectation value of the Lorentz gauge condition is satisfied: $\langle\Phi|\partial^\mu\hat{A}_\mu|\Phi\rangle=0$. Nevertheless, our formalism is also compatible with the Gupta-Bleuler method. In fact, substituting Eq. (52) into $\langle \Phi | \partial^{\mu} \hat{A}_{\mu} | \Phi \rangle = 0$, and consider Eq. (56) one can obtain

$$\langle \varPhi | \hat{b}(\boldsymbol{k},3) | \varPhi \rangle = -\mathrm{i} \langle \varPhi | [\hat{c}(\boldsymbol{k},0) - \hat{c}(\boldsymbol{k},3)] | \varPhi \rangle = 0 . \tag{60}$$

Eq. (60) also implies that $\hat{\boldsymbol{E}}_{\|}$ should be discarded.

In terms of the matrix forms of $\boldsymbol{\eta}(\boldsymbol{k},i)$ (i.e., $\boldsymbol{\varepsilon}(\boldsymbol{k},i)$), substituting the matrix forms of Eqs. (53) and (55) into Eq. (22), one can obtain the spinor field in the standard representation:

$$\hat{\psi}_{\mathrm{S}}(x) = \int \frac{\sqrt{\omega} \mathrm{d}^3 k}{\sqrt{2(2\pi)^3}} \sum_{i=1}^{2} \{ f(\boldsymbol{k},i) [\hat{b}(\boldsymbol{k},i) \exp(\mathrm{i}k \cdot x) + \hat{b}^{\dagger}(\boldsymbol{k},i) \exp(-\mathrm{i}k \cdot x)] \} , \tag{61}$$

where

$$f(\boldsymbol{k},1) = \frac{1}{\sqrt{2}} \begin{pmatrix} \boldsymbol{\varepsilon}(\boldsymbol{k},1) \\ \mathrm{i}\boldsymbol{\varepsilon}(\boldsymbol{k},2) \end{pmatrix} , \quad f(\boldsymbol{k},2) = \frac{1}{\sqrt{2}} \begin{pmatrix} \boldsymbol{\varepsilon}(\boldsymbol{k},2) \\ -\mathrm{i}\boldsymbol{\varepsilon}(\boldsymbol{k},1) \end{pmatrix} . \tag{62}$$

Using Eqs. (40), (49) and (62), one can obtain, for $i, j = 1, 2$,

$$f^{\dagger}(\boldsymbol{k},i) f(\boldsymbol{k},j) = \delta_{ij} , \quad f^{\dagger}(\boldsymbol{k},i) f(-\boldsymbol{k},j) = 0 , \tag{63-1}$$

$$f^{\dagger}(\boldsymbol{k},i) \boldsymbol{\alpha}_{\mathrm{S}} f(\boldsymbol{k},j) = \delta_{ij} \boldsymbol{k} / \omega , \quad f^{\dagger}(\boldsymbol{k},i) \boldsymbol{\alpha}_{\mathrm{S}} f(-\boldsymbol{k},j) = (0,0,0) . \tag{63-2}$$

Likewise, substituting the matrix forms of Eqs. (53) and (55) into Eq. (7), one can obtain the spinor field in the chiral representation:

$$\hat{\psi}_{\mathrm{C}}(x) = \int \frac{\sqrt{\omega} \mathrm{d}^3 k}{\sqrt{2(2\pi)^3}} \sum_{i=1}^{2} \{ g(\boldsymbol{k},i) [\hat{b}(\boldsymbol{k},i) \exp(\mathrm{i}k \cdot x) + \hat{b}^{\dagger}(\boldsymbol{k},i) \exp(-\mathrm{i}k \cdot x)] \} , \tag{64}$$

where

$$g(\boldsymbol{k},1) = \frac{1}{\sqrt{2}} \begin{pmatrix} \boldsymbol{e}_{1}(\boldsymbol{k}) \\ \boldsymbol{e}_{-1}(\boldsymbol{k}) \end{pmatrix} , \quad g(\boldsymbol{k},2) = \frac{1}{\sqrt{2}} \begin{pmatrix} -\mathrm{i}\boldsymbol{e}_{1}(\boldsymbol{k}) \\ \mathrm{i}\boldsymbol{e}_{-1}(\boldsymbol{k}) \end{pmatrix} . \tag{65}$$

Here we have defined the 3D circular polarization vectors as follows:

$$\boldsymbol{e}_1(\boldsymbol{k}) = \boldsymbol{e}^*_{-1}(\boldsymbol{k}) = \frac{\boldsymbol{\varepsilon}(\boldsymbol{k},1) + \mathrm{i}\boldsymbol{\varepsilon}(\boldsymbol{k},2)}{\sqrt{2}} = \frac{1}{\sqrt{2}|\boldsymbol{k}|}\begin{pmatrix} \frac{k_1 k_3 - \mathrm{i}k_2|\boldsymbol{k}|}{k_1 - \mathrm{i}k_2} \\ \frac{k_2 k_3 + \mathrm{i}k_1|\boldsymbol{k}|}{k_1 - \mathrm{i}k_2} \\ -(k_1 + \mathrm{i}k_2) \end{pmatrix}, \quad \boldsymbol{e}_0(\boldsymbol{k}) = \boldsymbol{\varepsilon}(\boldsymbol{k},3). \tag{66}$$

Using Eqs. (11) and (66) one can obtain the following formulae:

$$\boldsymbol{e}^\dagger_\lambda(\boldsymbol{k})\boldsymbol{e}_{\lambda'}(\boldsymbol{k}) = \delta_{\lambda\lambda'}, \quad \sum_\lambda \boldsymbol{e}_\lambda(\boldsymbol{k})\boldsymbol{e}^\dagger_\lambda(\boldsymbol{k}) = I_{3\times3}, \quad \lambda, \lambda' = \pm1, 0. \tag{67-1}$$

$$\boldsymbol{e}^\dagger_\lambda(-\boldsymbol{k})\boldsymbol{e}_\lambda(\boldsymbol{k}) = 0, \quad \boldsymbol{e}^\dagger_\lambda(\boldsymbol{k})\boldsymbol{\tau}\boldsymbol{e}_\lambda(-\boldsymbol{k}) = (0,0,0), \quad \lambda = \pm1, \tag{67-2}$$

$$\boldsymbol{e}^\dagger_1(\boldsymbol{k})\boldsymbol{\tau}\boldsymbol{e}_1(\boldsymbol{k}) = -\boldsymbol{e}^\dagger_{-1}(\boldsymbol{k})\boldsymbol{\tau}\boldsymbol{e}_{-1}(\boldsymbol{k}) = \boldsymbol{k}/\omega. \tag{67-3}$$

$$\boldsymbol{e}_1(-\boldsymbol{k}) = -\exp(\mathrm{i}\theta)\boldsymbol{e}_{-1}(\boldsymbol{k}), \quad \boldsymbol{e}_{-1}(-\boldsymbol{k}) = -\exp(-\mathrm{i}\theta)\boldsymbol{e}_1(\boldsymbol{k}), \tag{67-4}$$

where

$$\theta = \arctan[2k_1k_2/(k_1^2 - k_2^2)], \quad \exp(\mathrm{i}\theta) = (k_1 + \mathrm{i}k_2)/(k_1 - \mathrm{i}k_2). \tag{67-5}$$

Using Eqs. (65) and (67) one has, for $i = 1, 2$,

$$g^\dagger(\boldsymbol{k},i)g(\boldsymbol{k},j) = \delta_{ij}, \quad g^\dagger(\boldsymbol{k},i)\boldsymbol{\alpha}_{\mathrm{C}}g(\boldsymbol{k},j) = \delta_{ij}\boldsymbol{k}/\omega, \tag{68-1}$$

$$g^\dagger(\boldsymbol{k},i)g(-\boldsymbol{k},j) = 0, \quad g^\dagger(\boldsymbol{k},i)\boldsymbol{\alpha}_{\mathrm{C}}g(-\boldsymbol{k},j) = (0,0,0). \tag{68-2}$$

It is easy to show that

$$\frac{\boldsymbol{\tau}\cdot\boldsymbol{k}}{|\boldsymbol{k}|}\boldsymbol{e}_\lambda(\boldsymbol{k}) = \lambda\boldsymbol{e}_\lambda(\boldsymbol{k}), \quad \lambda = \pm1, 0. \tag{69}$$

Eq. (69) implies that $\boldsymbol{e}_0(\boldsymbol{k}) = \boldsymbol{\varepsilon}(\boldsymbol{k},3)$ (parallel to $\boldsymbol{k}$) denotes the longitudinal polarization vector, while $\boldsymbol{e}_{\pm1}(\boldsymbol{k})$ (perpendicular to $\boldsymbol{k}$) correspond to the right- and left-hand circular polarization vectors, respectively, and $\lambda = \pm1, 0$ represent the spin projections in the direction of $\boldsymbol{k}$ (i.e., $\lambda = \pm1$ represent the helicities of photons). Moreover, because of $(\boldsymbol{\tau}\cdot\boldsymbol{k})\boldsymbol{e}_0(\boldsymbol{k}) = (\boldsymbol{\tau}\cdot\boldsymbol{k})\boldsymbol{\varepsilon}(\boldsymbol{k},3) = 0$, substituting Eq. (50) and Eqs. (53)-(55) into Eq. (7) or (22), one can show that the dispersion relation $\omega_0^2 = 0$ in Eq. (44) corresponds to the one of the admixture of the longitudinal and scalar photons, and then for a radiation field the solution $\omega_0^2 = 0$ should be discarded. BTW, rotating $\boldsymbol{e}_{\pm1}(\boldsymbol{k})$ round the wave number vector $\boldsymbol{k}$, one can obtain another right- and left-hand circular polarization vectors, respectively. For the moment $\boldsymbol{e}_0(\boldsymbol{k}) = \boldsymbol{k}/|\boldsymbol{k}|$ is invariant. For example, one can use $\boldsymbol{e}'_{\pm1}(\boldsymbol{k}) = \exp(\pm \mathrm{i}\phi)\boldsymbol{e}_{\pm1}(\boldsymbol{k})$ instead of $\boldsymbol{e}_{\pm1}(\boldsymbol{k})$, where $\phi = -\arctan(k_2/k_1)$ and then $\exp(\mathrm{i}\phi) = -(k_1 - \mathrm{i}k_2)/(k_1^2 + k_2^2)$.

The annihilation and creation operators in Eq. (52) satisfy the commutation relations,

$$[\hat{c}(\boldsymbol{k}', s'), \hat{c}^\dagger(\boldsymbol{k}, s)] = \eta_{ss'}\delta^{(3)}(\boldsymbol{k} - \boldsymbol{k}'), \quad s, s' = 0,1,2,3, \tag{70}$$

where $\delta^{(3)}(\boldsymbol{k}' - \boldsymbol{k}) = \delta(k_1' - k_1)\delta(k_2' - k_2)\delta(k_3' - k_3)$, $\eta_{\mu\nu} = \mathrm{diag}(-1,1,1,1)$. Using Eqs. (56), and (70), one has

$$[\hat{b}(\boldsymbol{k}', i), \hat{b}^\dagger(\boldsymbol{k}, j)] = \delta_{ij}\delta^{(3)}(\boldsymbol{k} - \boldsymbol{k}'), \quad i, j = 1,2, \tag{71}$$

with the others vanishing. In particular, one has $[\hat{b}(\boldsymbol{k},3), \hat{b}^\dagger(\boldsymbol{k},3)] = 0$, which is due to the fact that the contributions of the longitudinal and scalar photons cancel each other.

Moreover, using Eqs. (7), (53) and (55) one can obtain,

$$\boldsymbol{F}_{\mathrm{R}}(x) = \int \frac{\sqrt{\omega}\mathrm{d}^3k}{\sqrt{2(2\pi)^3}} \boldsymbol{e}_1(\boldsymbol{k})[\hat{a}_1(\boldsymbol{k})\exp(\mathrm{i}k \cdot x) + \hat{a}_{-1}^\dagger(\boldsymbol{k})\exp(-\mathrm{i}k \cdot x)], \tag{72}$$

$$\boldsymbol{F}_{\mathrm{L}}(x) = \int \frac{\sqrt{\omega}\mathrm{d}^3k}{\sqrt{2(2\pi)^3}} \boldsymbol{e}_{-1}(\boldsymbol{k})[\hat{a}_{-1}(\boldsymbol{k})\exp(\mathrm{i}k \cdot x) + \hat{a}_1^\dagger(\boldsymbol{k})\exp(-\mathrm{i}k \cdot x)], \tag{73}$$

where

$$\hat{a}_{\pm1}(\boldsymbol{k}) = [\hat{b}(\boldsymbol{k},1) \mp \mathrm{i}\hat{b}(\boldsymbol{k},2)]/\sqrt{2}\,. \tag{74}$$

It is easy to show that

$$[\hat{a}_\lambda(\boldsymbol{k}), \hat{a}_{\lambda'}^\dagger(\boldsymbol{k}')] = \delta_{\lambda\lambda'}\delta^{(3)}(\boldsymbol{k} - \boldsymbol{k}'), \quad \lambda, \lambda' = \pm1. \tag{75}$$

It follows from Eqs. (69), (72) and (73) that $\boldsymbol{F}_{\mathrm{R}}(x)$ and $\boldsymbol{F}_{\mathrm{L}}(x)$ describe the transverse photons with the helicities of $\lambda = \pm1$, respectively. Each frequency component of $\boldsymbol{F}_{\mathrm{R}}(x)$ and $\boldsymbol{F}_{\mathrm{L}}(x)$ is the eigenstate of helicity, just as shown by Eq. (19). That is, the right-handed (left-handed) spinor having positive (negative) helicity. Moreover, taking $\boldsymbol{F}_{\mathrm{R}}(x)$ as an example, if its positive-frequency part has the momentum of $\boldsymbol{k}\in(-\infty,+\infty)$, then its negative-frequency part has the momentum of $-\boldsymbol{k}$. As a result, if the state vector of $\hat{a}_1^{\dagger}(\boldsymbol{k})|0\rangle$ has the spin of "up", then the one of $\hat{a}_{-1}^{\dagger}(\boldsymbol{k})|0\rangle$ has the spin of "down", but both of them have positive helicity (indicated by the right-hand circular polarization vector $\boldsymbol{e}_1(\boldsymbol{k})$ in $\boldsymbol{F}_{\mathrm{R}}(x)$).

As mentioned before, the spinors $\psi_{\mathrm{CR}}$ and $\psi_{\mathrm{CL}}$ defined by Eq. (20) are the eigenstates of the chirality operator with the eigenvalues $\pm 1$ respectively. They can be regarded as the 6-component versions of $\boldsymbol{F}_{\mathrm{R}}$ and $\boldsymbol{F}_{\mathrm{L}}$, respectively. From the point of view of quantized field, $\hat{N}_{\mathrm{circular}}(\boldsymbol{k},\lambda)=\hat{a}_{\lambda}^{\dagger}(\boldsymbol{k})\hat{a}_{\lambda}(\boldsymbol{k})$ ( $\lambda=\pm 1$ ) are the number operators of circular-polarized photons, while $\hat{N}(\boldsymbol{k},i)=\hat{b}^{\dagger}(\boldsymbol{k},i)\hat{b}(\boldsymbol{k},i)$ ( $i=1,2$ ) are the number operators of linear-polarized photons. One can show that $\sum_{\lambda}\hat{N}_{\mathrm{circular}}(\boldsymbol{k},\lambda)=\sum_{i}\hat{N}(\boldsymbol{k},i)$.

In our formalism, the 4D momentum of the electromagnetic field can be expressed as

$$J^{\mu}=\int\bar{\psi}(x)\beta^{\mu}\psi(x)\mathrm{d}^3x=\int((\boldsymbol{E}^2+\boldsymbol{H}^2)/2,\boldsymbol{E}\times\boldsymbol{H})\mathrm{d}^3x. \tag{76}$$

Substituting Eq. (61) (or Eq. (64)) into Eq. (78), using Eq. (63) (or (68)), and consider that

$$\int\frac{\mathrm{d}^3x}{(2\pi)^3}\exp[\pm\mathrm{i}(\boldsymbol{k}+\boldsymbol{k}')\cdot\boldsymbol{x}]=\delta^{(3)}(\boldsymbol{k}+\boldsymbol{k}'),\ \int\frac{\mathrm{d}^3x}{(2\pi)^3}\exp[\pm\mathrm{i}(\boldsymbol{k}-\boldsymbol{k}')\cdot\boldsymbol{x}]=\delta^{(3)}(\boldsymbol{k}-\boldsymbol{k}'), \tag{77}$$

one can obtain that, for $\hat{\psi}=\hat{\psi}_{\mathrm{S}},\hat{\psi}_{\mathrm{C}}$ and $\beta^{\mu}=\beta_{\mathrm{S}}^{\mu},\beta_{\mathrm{C}}^{\mu}$,

$$J^0=\int\hat{\bar{\psi}}(x)\beta^0\hat{\psi}(x)\mathrm{d}^3x=\int\mathrm{d}^3k\sum_{i=1}^{2}\omega[\hat{N}(\boldsymbol{k},i)+\delta^3(0)/2], \tag{78}$$

$$\boldsymbol{J}=\int\hat{\bar{\psi}}(x)\boldsymbol{\beta}\hat{\psi}(x)\mathrm{d}^3x=\int\mathrm{d}^3k\sum_{i=1}^{2}\boldsymbol{k}\hat{N}(\boldsymbol{k},i), \tag{79}$$

where $\hat{N}(\boldsymbol{k},i)=\hat{b}^{\dagger}(\boldsymbol{k},i)\hat{b}(\boldsymbol{k},i)$ ( $i=1,2$ ) represents the number operator of the photon field, it is the number operator of linear-polarized photons.

BTW, with different notations and conventions, we have presented a theoretical model with the duality between the electric and magnetic fields, where all the $\lambda = \pm 1, 0$ photons are taken into account, from which we have shown that the photonic zitterbewegung occurs in the presence of virtual longitudinal and scalar photons, and the gravitational interaction may result in the zitterbewegung of photons [37].

## 4. Some symmetries in terms of the (1, 0) ⊕ (0, 1) spinor description

Let us introduce a Lorentz scalar as follows

$$\mathcal{L} = \bar{\psi}(x)(\mathrm{i}\beta^{\mu}\partial_{\mu})\psi(x)\,. \tag{80}$$

We call $\mathcal{L}$ the pseudo-Lagrangian density, and call $S = \int \mathcal{L}\mathrm{d}^4 x$ the pseudo-action. Applying the principle of least pseudo-action in $S = \int \mathcal{L}\mathrm{d}^4 x$ one can easily obtain the Dirac-like equation (42), but notice that the dimension of $\mathcal{L}$ is [1/length]$^5$ instead of [1/length]$^4$, which implies that $S = \int \mathcal{L}\mathrm{d}^4 x$ has the dimension of [1/length]$^1$ rather than [1/length]$^0$. In our formalism, all the symmetries of the Dirac-like equation are equivalent to the ones of $\mathcal{L}$ given by Eq. (80). Therefore, let us discuss some symmetries of $\mathcal{L}$.

1) *Lorentz invariance*

According to Section 2, the Lorentz invariance of the Dirac-like equation (42) is self-evident. Nonetheless, here we will discuss it in another way. Under a proper Lorentz transformation of $x^{\mu} \to x'^{\mu} = \Lambda^{\mu\nu}x_{\nu}$, the photon field $\psi = \psi_{\mathrm{S}}$ (or $\psi_{\mathrm{C}}$) transforms linearly in the way (here for the sake of character display, let $L^{+}$, instead of $L^{\dagger}$, denote the Hermitian adjoint of $L$):

$$\psi(x) \to \psi'(x') = L(\Lambda)\psi(x)\,,\quad \bar{\psi}(x) \to \bar{\psi}'(x') = \bar{\psi}(x)\beta^{0}L^{+}(\Lambda)\beta^{0}\,. \tag{81}$$

According to Section 2, we have ( $\mu, \nu = 01, 2, 3$ , $l, m, n = 1, 2, 3$ )

$$L(\Lambda) = \exp(-\mathrm{i}\omega_{\mu\nu} S^{\mu\nu}/2) = \exp(\mathrm{i}\boldsymbol{\theta} \cdot \boldsymbol{\Sigma} + \boldsymbol{\varsigma} \cdot \boldsymbol{\alpha}), \tag{82}$$

where $\omega_{0l} = \varsigma_l$, $\omega_{lm} = -\varepsilon_{lmn}\theta_n$, while $S_{lm} = \varepsilon_{lmn}\Sigma^n$ and $S^{0l} = \mathrm{i}\alpha^l$ are determined by Eq. (40) or Eq. (41). Under an infinitesimal Lorentz transformation with $\omega^{\mu\nu} \to 0$, one has

$$\Lambda^{\mu\nu} = \eta^{\mu\nu} - \omega^{\mu\nu}, \quad L(\Lambda) = 1 - \mathrm{i}\omega_{\mu\nu} S^{\mu\nu}/2. \tag{83}$$

For the moment, the pseudo-Lagrangian density transforms as

$$\mathcal{L} = \bar{\psi}(\mathrm{i}\beta^{\mu}\partial_{\mu})\psi \to \mathcal{L}' = \bar{\psi}\beta^0 L^{+}\beta^0[\mathrm{i}\beta^{\mu}(\partial_{\mu} - \omega_{\mu\nu}\partial^{\nu})]L\psi . \tag{84}$$

Let $\delta\mathcal{L} \equiv \mathcal{L}' - \mathcal{L}$, only keep the first-order infinitesimal, on can obtain

$$\delta\mathcal{L} = \omega^{\mu\nu}\Delta_{\mu\nu}/2 = (\omega^{\mu\nu}/2)\bar{\psi}[\mathrm{i}(\beta_{\nu}\partial_{\mu} - \beta_{\mu}\partial_{\nu}) + (\beta^{\rho}S_{\mu\nu} - S_{\mu\nu}\beta^{\rho})\partial_{\rho}]\psi , \tag{85}$$

where

$$\Delta_{\mu\nu} = -\Delta_{\nu\mu} = \bar{\psi}[\mathrm{i}(\beta_{\nu}\partial_{\mu} - \beta_{\mu}\partial_{\nu}) + (\beta^{\rho}S_{\mu\nu} - S_{\mu\nu}\beta^{\rho})\partial_{\rho}]\psi . \tag{86}$$

A necessary and sufficient condition of $\mathcal{L}$ being Lorentz invariant is $\delta\mathcal{L} = 0$, that is, $\Delta_{\mu\nu} = 0$. Based on Eq. (86) one can discuss as follows:

(1) As $\mu = \nu$, $\Delta_{\mu\nu} = 0$, then $\delta\mathcal{L} = 0$ is valid;

(2) As $\mu = l$, $\nu = m$ ($l, m = 1, 2, 3$), using $S_{lm} = \varepsilon_{lmn}\Sigma^n$, $\beta^l\Sigma^m - \Sigma^m\beta^l = \mathrm{i}\varepsilon^{lmn}\beta_n$, and $\beta^0\boldsymbol{\Sigma} = \boldsymbol{\Sigma}\beta^0$, it follows from Eq. (86) that

$$\mathrm{i}(\beta_m\partial_l - \beta_l\partial_m) + \varepsilon_{lmn}(\beta^{\rho}\Sigma^n - \Sigma^n\beta^{\rho})\partial_{\rho} = 0, \tag{87}$$

which implies that $\Delta_{lm} = 0$, and then $\delta\mathcal{L} = 0$ is valid;

(3) As $\mu = l = 1, 2, 3$, $\nu = 0$, using $S_{l0} = -S_{0l} = \mathrm{i}\alpha^l$, $\beta_0 = -\beta^0$, from Eq. (86) one has

$$\Delta_{l0} = -\Delta_{0l} = \bar{\psi}[\mathrm{i}(\beta_0\partial_l - \beta_l\partial_0) + \mathrm{i}(\beta^{\rho}\alpha^l - \alpha^l\beta^{\rho})\partial_{\rho}]\psi , \tag{88}$$

using $\mathrm{i}\beta^{\rho}\partial_{\rho}\psi = 0$, $\boldsymbol{\alpha} = \beta^0\boldsymbol{\beta} = -\boldsymbol{\beta}\beta^0$, one has

$$\Delta_{l0} = -\Delta_{0l} = \mathrm{i}\,\bar{\psi}\beta^0[-\partial_l + \alpha^m\alpha^l\partial_m]\psi . \tag{89}$$

Using Eq. (40) or Eq. (41) and Eq. (11), $\nabla = (\partial_1, \partial_2, \partial_3)$, it is easy to show that

$$\Delta_{l0}=-\Delta_{0l}=-\mathrm{i}(E_l\nabla\cdot\boldsymbol{E}+H_l\nabla\cdot\boldsymbol{H})/2=0\,, \tag{90}$$

and then $\delta\mathcal{L}=0$ is valid.

The statements (1), (2) and (3) exhaust all cases, therefore, the pseudo-Lagrangian density $\mathcal{L}$ is Lorentz invariant, which implies that the Dirac-like equation (42) is Lorentz covariant. In fact, using Eq. (6-2) the covariance of Eq. (42) can also be proven directly.

By the way, Eq. (86) implies that,

$$\Delta_{\mu\nu}-(\Delta_{\mu\nu})^{\dagger}=\mathrm{i}[\partial_{\mu}(\bar{\psi}\beta_{\nu}\psi)-\partial_{\nu}(\bar{\psi}\beta_{\mu}\psi)]+\partial_{\rho}[\bar{\psi}(\beta^{\rho}S_{\mu\nu}-S_{\mu\nu}\beta^{\rho})\psi]=0\,. \tag{91}$$

Let

$$G_{\mu\nu}=\partial_{\mu}(\bar{\psi}\beta_{\nu}\psi)-\partial_{\nu}(\bar{\psi}\beta_{\mu}\psi)\,,\quad Q^{\rho}_{\mu\nu}=\bar{\psi}(\beta^{\rho}S_{\mu\nu}-S_{\mu\nu}\beta^{\rho})\psi\,. \tag{92}$$

Obviously, $G_{\mu\nu}$ is the 4D curl of $\bar{\psi}\beta_{\mu}\psi$, and Eq. (91) implies that,

$$\partial_{\rho}Q^{\rho}_{\mu\nu}=-\mathrm{i}G_{\mu\nu}\,. \tag{93}$$

Later we will show that $\bar{\psi}\beta_{\mu}\psi$ is the energy and momentum densities of the photon field.

2) *Generalized gauge symmetry*

Obviously, under the following transformation (generalized gauge transformation, say)

$$\psi(x)\to\exp(-\mathrm{i}\theta)\psi(x)\,, \tag{94}$$

the pseudo-Lagrangian density (80) (and then the Dirac-like equation (42)) is invariant, where $\theta$ is a real constant. The corresponding conserved current (rather than current *density*) $J^{\mu}$ is a 4D vector:

$$J^{\mu}=\int\bar{\psi}(x)\beta^{\mu}\psi(x)\mathrm{d}^3x=\int((\boldsymbol{E}^2+\boldsymbol{H}^2)/2,\boldsymbol{E}\times\boldsymbol{H})\mathrm{d}^3x\,. \tag{95}$$

It is the 4D momentum of the photon field (playing the role of a gravitational charge carried by the photon field). The 4D current vector satisfies the integral form of Poynting's theorem, i.e., $\partial_{\mu}J^{\mu}=0$, it is the continuity equation of the photon field.

3) *Space inversion symmetry*

Under space inversion (parity operation) of $\boldsymbol{x}\to-\boldsymbol{x}$ (say, P) , one can prove that the pseudo-Lagrangian density (80) (and then the Dirac-like equation (42)) is invariant, where the field quantity $\psi(x)$ transforms as

$$\psi(t,\boldsymbol{x})\to \mathrm{P}\psi(t,\boldsymbol{x})\mathrm{P}^{-1}=\beta^{0}\psi(t,-\boldsymbol{x}). \tag{96}$$

To study Eq. (96), let us rotate $\boldsymbol{e}_{\pm1}(\boldsymbol{k})$ round the wave number vector $\boldsymbol{k}$ through the angle $\phi=-\arctan(k_2/k_1)$ (and then $\exp(\mathrm{i}\phi)=-(k_1-\mathrm{i}k_2)/(k_1^2+k_2^2)$), in such a way we obtain another right- and left-hand circular polarization vectors:

$$\boldsymbol{e}_{\pm1}(\boldsymbol{k})\to\boldsymbol{e}'_{\pm1}(\boldsymbol{k})=\exp(\pm\mathrm{i}\phi)\boldsymbol{e}_{\pm1}(\boldsymbol{k}),\ \ \phi=-\arctan(k_2/k_1). \tag{97}$$

For the moment, Eq. (67) is also valid for $\boldsymbol{e}'_{\pm1}(\boldsymbol{k})$, but for Eq. (67-4) being replaced with

$$\boldsymbol{e}'_{\pm1}(-\boldsymbol{k})=\boldsymbol{e}'_{\mp}(\boldsymbol{k}). \tag{98}$$

As an example, let us consider the spinor field in the chiral representation given by Eq. (64), but for $g(\boldsymbol{k},i)$ being defined by $\boldsymbol{e}'_{\pm1}(\boldsymbol{k})$ instead of $\boldsymbol{e}_{\pm1}(\boldsymbol{k})$:

$$g(\boldsymbol{k},1)=\frac{1}{\sqrt{2}}\begin{pmatrix}\boldsymbol{e}'_{1}(\boldsymbol{k})\\ \boldsymbol{e}'_{-1}(\boldsymbol{k})\end{pmatrix},\ \ g(\boldsymbol{k},2)=\frac{1}{\sqrt{2}}\begin{pmatrix}-\mathrm{i}\boldsymbol{e}'_{1}(\boldsymbol{k})\\ \mathrm{i}\boldsymbol{e}'_{-1}(\boldsymbol{k})\end{pmatrix}. \tag{99}$$

Using Eqs. (41), (98) and (99), one has

$$\beta_{\mathrm{C}}^{0}g(-\boldsymbol{k},i)=g(\boldsymbol{k},i),\ \ i=1,2. \tag{100}$$

Substituting Eqs. (64) and (99) into Eq. (96), using Eq. (100) one can obtain

$$\mathrm{P}\hat{b}(\boldsymbol{k},1)\mathrm{P}^{-1}=\hat{b}(-\boldsymbol{k},1),\ \ \mathrm{P}\hat{b}(\boldsymbol{k},2)\mathrm{P}^{-1}=-\hat{b}(-\boldsymbol{k},2). \tag{101}$$

4) *Time reversal symmetry*

Under the time reversal transformation of $t\to-t$ (denoted by T), one can show that the pseudo-Lagrangian density Eq. (80) (and then the Dirac-like equation (42)) is invariant, and the field quantity $\psi(x)$ transforms as

$$\psi(t,\boldsymbol{x}) \to \mathrm{T}\,\psi(\boldsymbol{x},t)\mathrm{T}^{-1} = \psi(\boldsymbol{x},-t)\,. \qquad (102)$$

Using $\mathrm{TiT}^{-1} = -\mathrm{i}$ one has

$$\mathrm{T}\boldsymbol{e}_{\pm}(\boldsymbol{k})\mathrm{T}^{-1} = \boldsymbol{e}_{\mp}(\boldsymbol{k}),\quad \mathrm{T}\boldsymbol{e}'_{\pm 1}(\boldsymbol{k})\mathrm{T}^{-1} = \boldsymbol{e}'_{\mp}(\boldsymbol{k})\,. \qquad (103)$$

Using $\mathrm{TiT}^{-1} = -\mathrm{i}$, Eqs. (102) and (103) one can prove that

$$\mathrm{T}\hat{b}(\boldsymbol{k},1)\mathrm{T}^{-1} = \hat{b}(-\boldsymbol{k},1),\quad \mathrm{T}\hat{b}(\boldsymbol{k},2)\mathrm{T}^{-1} = -\hat{b}(-\boldsymbol{k},2)\,. \qquad (104)$$

5) *Charge-conjugation transformation*

Using Eq. (40) or Eq. (41) one can easily show that, under the transformation

$$\psi \to \psi' = \mathrm{C}\,\psi\mathrm{C}^{-1} \equiv \beta^{0}\psi^{*}, \qquad (105)$$

one has $\psi' = \psi$, and then the pseudo-Lagrangian density Eq. (80) (as well as the Dirac-like equation (42)) is invariant. This transformation is called the charge-conjugation transformation, and the fact of $\psi' = \psi$ implies that the photon field is a neutral field.

6) *Chiral transformation*

If a $n \times n$ self-adjoint matrix $A = A^{\dagger}$ satisfies the conditions of $A^{2l} = I_{n\times n}$ and $A^{2l+1} = A$, $l = 0,1,2...$, one has (the unit matrix $I_{n\times n}$ is left out provided that it is equivalent to 1, the same later)

$$\exp(\pm\mathrm{i}A\theta) = \cos\theta \pm \mathrm{i}A\sin\theta\,. \qquad (106)$$

Using Eq. (106) we can define the chiral transformation as follows ($\theta$ is a constant):

$$\psi \to \psi' = (\cos\theta + \mathrm{i}\beta^{5}\sin\theta)\psi = \exp(\mathrm{i}\beta^{5}\theta)\psi\,. \qquad (107)$$

Under the transformation of Eq. (107), using Eq. (40) or Eq. (41) one can show that the pseudo-Lagrangian density (80) (and then the Dirac-like equation (42)) is invariant. The chiral transformation of Eq. (107) can be rewritten as

$$\boldsymbol{E} \to \boldsymbol{E}' = \boldsymbol{E}\cos\theta - \boldsymbol{H}\sin\theta\,,\quad \boldsymbol{H} \to \boldsymbol{H}' = \boldsymbol{H}\cos\theta + \boldsymbol{E}\sin\theta\,. \qquad (108)$$

For such a transformation, D. R. Brill and J. A. Wheeler have presented an interpretation [63]: if $\boldsymbol{E}$ and $\boldsymbol{H}$ represent the principal polarization directions of a monochromatic plane wave, the transformation rotates the axes of polarization by the angle $\theta$. However, the authors in Ref. [44] have presented another interpretation: in the absence of sources, the splitting of the electromagnetic field into the electric and magnetic parts is not absolute. BTW, it follows from the invariance of Eq. (80) that $\partial_{\mu} j^{\mu 5}=0$. However, it is trivial because of $j^{\mu 5}=\bar{\psi}(x)\beta^{\mu}\beta^{5}\psi(x)=0$.

7) *Transformation under the operation of the chirality operator* $\beta^5$

We have defined 6-component right- and left-handed spinors (in chiral representation) via Eq. (20), we here restate it in a more general manner. Let $\psi=\psi_{\mathrm{S}},\psi_{\mathrm{C}}$, $\psi_{\mathrm{R}}=\psi_{\mathrm{SR}},\psi_{\mathrm{CR}}$, $\psi_{\mathrm{L}}=\psi_{\mathrm{SL}},\psi_{\mathrm{CL}}$, $\beta^5=\beta^5_{\mathrm{S}},\beta^5_{\mathrm{C}}$, using Eq. (40) or (41) let us define 6-component right- and left-handed spinors as follows, respectively,

$$\psi_{\mathrm{R}}=\frac{1}{2}(1+\beta^5)\psi\,,\quad \psi_{\mathrm{L}}=\frac{1}{2}(1-\beta^5)\psi\,. \tag{109}$$

In standard representation, Eq. (109) becomes

$$\psi_{\mathrm{SR}}=\frac{1}{2}(1+\beta^5_{\mathrm{S}})\psi_{\mathrm{S}}=\frac{1}{\sqrt{2}}\begin{pmatrix}\boldsymbol{F}_{\mathrm{R}}\\ \boldsymbol{F}_{\mathrm{R}}\end{pmatrix},\quad \psi_{\mathrm{SL}}=\frac{1}{2}(1-\beta^5_{\mathrm{S}})\psi_{\mathrm{S}}=\frac{1}{\sqrt{2}}\begin{pmatrix}\boldsymbol{F}_{\mathrm{L}}\\ -\boldsymbol{F}_{\mathrm{L}}\end{pmatrix}. \tag{110}$$

In chiral representation, Eq. (109) becomes

$$\psi_{\mathrm{CR}}=\frac{1}{2}(1+\beta^5_{\mathrm{C}})\psi_{\mathrm{C}}=\begin{pmatrix}\boldsymbol{F}_{\mathrm{R}}\\ 0\end{pmatrix},\quad \psi_{\mathrm{CL}}=\frac{1}{2}(1-\beta^5_{\mathrm{C}})\psi_{\mathrm{C}}=\begin{pmatrix}0\\ \boldsymbol{F}_{\mathrm{L}}\end{pmatrix}. \tag{111}$$

It is easy to prove that

$$\beta^5\psi_{\mathrm{R}}=\psi_{\mathrm{R}}\,,\quad \beta^5\psi_{\mathrm{L}}=-\psi_{\mathrm{L}}\,, \tag{112-1}$$

$$\mathcal{L}=\mathcal{L}_{\mathrm{R}}+\mathcal{L}_{\mathrm{L}}\,,\quad \mathcal{L}_{\mathrm{R}}=\bar{\psi}_{\mathrm{R}}(\mathrm{i}\beta^{\mu}\partial_{\mu})\psi_{\mathrm{R}}\,,\quad \mathcal{L}_{\mathrm{L}}=\bar{\psi}_{\mathrm{L}}(\mathrm{i}\beta^{\mu}\partial_{\mu})\psi_{\mathrm{L}}\,, \tag{112-2}$$

$$\psi\to\psi'=\beta^5\psi\,,\quad \mathcal{L}\to\mathcal{L}'=\mathcal{L}\,. \tag{112-3}$$

Eq. (112-3) implies that under the operation of the chirality operator, the pseudo-Lagrangian density (80) (and then the Dirac-like equation (42)) is invariant.

8) *The angular momentum of the photon field*

According to electrodynamics, the total angular momentum of electromagnetic field is

$$\boldsymbol{J}=\int[\boldsymbol{x}\times(\boldsymbol{E}\times\boldsymbol{H})]\mathrm{d}^3x=\boldsymbol{L}+\boldsymbol{S}\,, \tag{113}$$

where $\boldsymbol{L}$ and $\boldsymbol{S}$ are the orbital and spin angular momentums, respectively, i.e.,

$$\boldsymbol{L}=\int[E_i(\boldsymbol{x}\times\nabla)A_i]\mathrm{d}^3x\,,\quad \boldsymbol{S}=\int(\boldsymbol{E}\times\boldsymbol{A})\mathrm{d}^3x\,. \tag{114}$$

In terms of the $(1,0)\oplus(0,1)$ spinor, the total angular momentum of the photon field can be expressed as

$$\boldsymbol{J}=\boldsymbol{L}+\boldsymbol{S}=\int\psi^{\dagger}(\boldsymbol{x}\times\boldsymbol{\alpha})\psi\mathrm{d}^3x=\int\bar{\psi}(\boldsymbol{x}\times\boldsymbol{\beta})\psi\mathrm{d}^3x\,. \tag{115}$$

Similar discussions have first been presented by S. M. Barnett [47]. Eq. (115) can be regarded as the pure spatial components of the quantity:

$$J^{\mu\nu}=\int\bar{\psi}(x^{\mu}\beta^{\nu}-x^{\nu}\beta^{\mu})\psi\mathrm{d}^3x\,. \tag{116}$$

For our purpose, let us present a useful formula ($i,j=1,2,3$):

$$T_{ij}=\psi^{\dagger}(2\alpha_i\alpha_j-\delta_{ij})\psi=T_{ji}^{\dagger}=\delta_{ij}(\boldsymbol{E}^{\dagger}\boldsymbol{E}+\boldsymbol{H}^{\dagger}\boldsymbol{H})/2-(E_j^{\dagger}E_i+H_j^{\dagger}H_i)\,. \tag{117}$$

As for the angular momentum of the photon field, a new and interesting study based on the $(1,0)\oplus(0,1)$ spinor will be presented in our next work.

## 5. The Dirac-like equation in a linear medium

In this Section, let us temporarily apply the SI units of measurement (let $c=1/\sqrt{\varepsilon_0\mu_0}$ denote the velocity of light in Minkowski vacuum, $\hbar$ denote the Planck constant divided by $2\pi$). Let the free charge and current densities vanish, the Maxwell equations in a linear medium can be written as

$$\nabla\times\boldsymbol{H}-\partial\boldsymbol{D}/\partial t=0,\ \ \nabla\times\boldsymbol{E}+\partial\boldsymbol{B}/\partial t=0, \tag{118}$$

$$\nabla\cdot\boldsymbol{D}=0,\ \ \nabla\cdot\boldsymbol{B}=0. \tag{119}$$

Likewise, the constraint conditions (Eq. (119)) are actually contained by the dynamical equations (Eq. (118)). For the linear medium, the constitutive relations can be written as

$$\boldsymbol{D}=\varepsilon_0\varepsilon_r\boldsymbol{E}=\varepsilon\boldsymbol{E},\ \ \boldsymbol{B}=\mu_0\mu_r\boldsymbol{H}=\mu\boldsymbol{H}, \tag{120}$$

where $\varepsilon=\varepsilon_0\varepsilon_r$, $\mu=\mu_0\mu_r$, $\varepsilon_r$ and $\mu_r$ are the relative dielectric constant and relative permeability, respectively, they are two scalar functions. Let $n=\sqrt{\varepsilon_r\mu_r}$ be the refractive index of the linear medium, and then $u=1/\sqrt{\varepsilon\mu}=c/n$ is the velocity of light in the linear medium. It follows from Eqs. (118) and (119) that

$$\begin{cases}[\nabla-(\nabla\ln\sqrt{\mu_r})]\times(\sqrt{\mu}\boldsymbol{H})=(\frac{n}{c}\frac{\partial}{\partial t}+\frac{n}{c}\frac{\partial\ln\sqrt{\varepsilon_r}}{\partial t})(\sqrt{\varepsilon}\boldsymbol{E})\\ [\nabla-(\nabla\ln\sqrt{\varepsilon_r})]\times(\sqrt{\varepsilon}\boldsymbol{E})=-(\frac{n}{c}\frac{\partial}{\partial t}+\frac{n}{c}\frac{\partial\ln\sqrt{\mu_r}}{\partial t})(\sqrt{\mu}\boldsymbol{H})\\ [\nabla+(\nabla\ln\sqrt{\varepsilon_r})]\cdot(\sqrt{\varepsilon}\boldsymbol{E})=0\\ [\nabla+(\nabla\ln\sqrt{\mu_r})]\cdot(\sqrt{\mu}\boldsymbol{H})=0\end{cases}. \tag{121}$$

As mentioned before, the column-matrix forms of the vectors $\boldsymbol{E}=(E_1,E_2,E_3)$ and $\boldsymbol{H}=(H_1,H_2,H_3)$ are also denoted as $\boldsymbol{E}$ and $\boldsymbol{H}$, i.e., $\boldsymbol{E}=\begin{pmatrix}E_1 & E_2 & E_3\end{pmatrix}^{\mathrm{T}}$, $\boldsymbol{H}=\begin{pmatrix}H_1 & H_2 & H_3\end{pmatrix}^{\mathrm{T}}$. For the moment, the photon wave function $\psi=\psi_{\mathrm{S}},\psi_{\mathrm{C}}$ becomes

$$\psi_{\mathrm{S}}=\frac{1}{\sqrt{2}}\begin{pmatrix}\sqrt{\varepsilon}\boldsymbol{E}\\ \mathrm{i}\sqrt{\mu}\boldsymbol{H}\end{pmatrix},\ \ \psi_{\mathrm{C}}=\frac{1}{2}\begin{pmatrix}\sqrt{\varepsilon}\boldsymbol{E}+\mathrm{i}\sqrt{\mu}\boldsymbol{H}\\ \sqrt{\varepsilon}\boldsymbol{E}-\mathrm{i}\sqrt{\mu}\boldsymbol{H}\end{pmatrix}. \tag{122}$$

Let us define ($\nu=0,1,2,3$)

$$\partial_\nu=(n\partial/c\partial t,\nabla),\ \ \chi^\nu=\partial_\nu\ln\sqrt{\varepsilon_r}=(\chi^0,\boldsymbol{\chi}),\ \ \eta^\nu=\partial_\nu\ln\sqrt{\mu_r}=(\eta^0,\boldsymbol{\eta}), \tag{123}$$

where $\chi^0=-\chi_0$, $\chi^i=\chi_i$, $\eta^0=-\eta_0$, $\eta^i=\eta_i$, $i=1,2,3$. Obviously, $\boldsymbol{\chi}=\nabla\ln\sqrt{\varepsilon_r}$ and $\boldsymbol{\eta}=\nabla\ln\sqrt{\mu_r}$. It follows from $n=\sqrt{\varepsilon_r\mu_r}$ that $\nabla\ln n=\boldsymbol{\chi}+\boldsymbol{\eta}$. Let

$$\varphi_\nu = (\varphi_0, \boldsymbol{\varphi}) = \begin{pmatrix} \chi_\nu I_{3\times3} & 0 \\ 0 & \eta_\nu I_{3\times3} \end{pmatrix}, \ \varphi^0 = -\varphi_0, \ \varphi^i = \varphi_i, \ i = 1,2,3. \quad (124)$$

Using Eqs. (121)-(124), one can obtain the Dirac-like equation in the linear medium,

$$\mathrm{i}\hbar\beta^\nu(\partial_\nu - \varphi_\nu)\psi(x) = 0, \text{ or } \mathrm{i}\hbar\beta^\nu D_\nu\psi(x) = 0, \quad (125)$$

where $D_\nu = \partial_\nu - \varphi_\nu = (D_0, \boldsymbol{D})$, $\nu = 0,1,2,3$, $\psi = \psi_{\mathrm{S}}, \psi_{\mathrm{C}}$ is given by Eq. (122), and $\beta^\nu = \beta^\nu_{\mathrm{S}}, \beta^\nu_{\mathrm{C}}$ is given by Eqs. (40) and (41). For the moment, the constraint conditions associated with the dynamical equation (125) are

$$(\nabla + \boldsymbol{\chi}) \cdot (\sqrt{\varepsilon}\boldsymbol{E}) = 0, \ (\nabla + \boldsymbol{\eta}) \cdot (\sqrt{\mu}\boldsymbol{H}) = 0. \quad (126)$$

Let $f$ be a function, one has

$$(\nabla\partial_0 - \partial_0\nabla) f = (\nabla \ln n)\partial_0 f, \quad (127)$$

$$(\partial_0\boldsymbol{\varphi} - \nabla\varphi_0) f = [2(\partial_0\boldsymbol{\varphi}) - \varphi_0\nabla + \boldsymbol{\varphi}\partial_0 - (\nabla \ln n)\varphi_0] f. \quad (128)$$

Using Eqs. (38), (39), (127) and (128) one has

$$\boldsymbol{D}D_0 - D_0\boldsymbol{D} = (\nabla \ln n)\partial_0 + 2(\partial_0\boldsymbol{\varphi}) - (\nabla \ln n)\varphi_0, \quad (129)$$

$$(\boldsymbol{D}\cdot\boldsymbol{\beta})(\boldsymbol{\beta}\cdot\boldsymbol{D}) = -\nabla^2 - \boldsymbol{\varphi}\cdot\boldsymbol{\varphi} + \boldsymbol{\varphi}\cdot\nabla + \nabla\cdot\boldsymbol{\varphi} + 2\mathrm{i}\boldsymbol{\Sigma}\cdot(\boldsymbol{\varphi}\times\nabla) + \Omega, \quad (130)$$

$$(D_\mu\beta^\mu)(\beta^\nu D_\nu) = D_0D_0 + (\boldsymbol{D}\cdot\boldsymbol{\beta})(\boldsymbol{\beta}\cdot\boldsymbol{D}) - \boldsymbol{\alpha}\cdot(\boldsymbol{D}D_0 - D_0\boldsymbol{D}) + (\chi_\mu - \eta_\mu)\beta^\mu D_0, \quad (131)$$

where $\boldsymbol{\alpha} = \boldsymbol{\alpha}_{\mathrm{S}}, \boldsymbol{\alpha}_{\mathrm{C}}$ and $\boldsymbol{\Sigma} = \boldsymbol{\Sigma}_{\mathrm{S}} = \boldsymbol{\Sigma}_{\mathrm{C}}$ are given by Eqs. (40) and (41), and

$$\Omega = \begin{pmatrix} \boldsymbol{X}\boldsymbol{X}^{\mathrm{T}} & 0 \\ 0 & \boldsymbol{Y}\boldsymbol{Y}^{\mathrm{T}} \end{pmatrix}, \ \boldsymbol{X} = \begin{pmatrix} \partial_1 - \chi_1 \\ \partial_2 - \chi_2 \\ \partial_3 - \chi_3 \end{pmatrix}, \ \boldsymbol{Y} = \begin{pmatrix} \partial_1 - \eta_1 \\ \partial_2 - \eta_2 \\ \partial_3 - \eta_3 \end{pmatrix}. \quad (132)$$

Using Eqs. (127)- (131), one has

$$\begin{aligned}(D_\mu\beta^\mu)(\beta^\nu D_\nu) = \partial^\mu\partial_\mu - \varphi^\mu\varphi_\mu + \varphi^\mu\partial_\mu - \partial^\mu\varphi_\mu + (\chi_\mu - \eta_\mu)\beta^\mu D_0 + \Omega \\ + 2\mathrm{i}\boldsymbol{\Sigma}\cdot(\boldsymbol{\varphi}\times\nabla) - \boldsymbol{\alpha}\cdot[(\nabla \ln n)\partial_0 + 2(\partial_0\boldsymbol{\varphi}) - (\nabla \ln n)\varphi_0]\end{aligned}. \quad (133)$$

As discussed before, the 4D spin tensor $S_{\mu\nu}$ of the photon field consists of the components $S_{lm} = \varepsilon_{lmn}\Sigma_n$ and $S_{0l} = -\mathrm{i}\alpha_l$. Therefore, the terms related to the 4D spin tensor is separated out in the second line on the right-hand side of Eq. (133).

## 6. Preliminary investigation on the spin-orbit interaction of the photon field

Now, let us come back to the natural units $\hbar = c = 1$, and take $\varepsilon_0 = \mu_0 = 1$. Using Eqs. (125) and (133), one has

$$\begin{aligned}&[\partial^\mu\partial_\mu - \varphi^\mu\varphi_\mu + \varphi^\mu\partial_\mu - \partial^\mu\varphi_\mu + (\chi_\mu - \eta_\mu)\beta^\mu D_0 + \Omega]\psi(x)\\ &+\{2\mathrm{i}\boldsymbol{\Sigma}\cdot(\boldsymbol{\varphi}\times\nabla) - \boldsymbol{\alpha}\cdot[(\nabla\ln n)\partial_0 + 2(\partial_0\boldsymbol{\varphi}) - (\nabla\ln n)\varphi_0]\}\psi(x) = 0\end{aligned}. \tag{134}$$

The second line on the left-hand side of Eq. (134) is related to the 4D spin tensor of the photon field, and then it represents the contributions from the spin-orbit interactions. It follows from Eqs. (123) and (124) that, the contributions from the spin-orbit interactions become vanishing if $\nabla\varepsilon_r = \nabla\mu_r = 0$, which implies that the spin-orbit interactions of the photon field only exist in an inhomogeneous medium.

Note that in Eq. (123) we have defined $\partial_\mu = (n\partial_t, \nabla)$ (rather than $\partial_\mu = (\partial_t, \nabla)$), where $\partial_t = \partial/\partial t$. As an example, let $\chi_0 = \eta_0 = 0$ and $\boldsymbol{\chi} = \boldsymbol{\eta}$, and then $\varphi_0 = 0$, $\boldsymbol{\varphi} = \boldsymbol{\chi} \boldsymbol{I}_{6\times6} = \boldsymbol{\chi}$, $\nabla\ln n = 2\boldsymbol{\chi}$, and $\Omega = I_{2\times2}\otimes \boldsymbol{X}\boldsymbol{X}^{\mathrm{T}}$. Eq. (134) becomes ( $\chi^2 = \boldsymbol{\chi}\cdot\boldsymbol{\chi}$ )

$$[-n^2\partial_t^2 + \nabla^2 - \chi^2 + \boldsymbol{\chi}\cdot\nabla - (\nabla\cdot\boldsymbol{\chi}) + \Omega]\psi(x) + [2\mathrm{i}\boldsymbol{\Sigma}\cdot(\boldsymbol{\chi}\times\nabla) - 2(\boldsymbol{\alpha}\cdot\boldsymbol{\chi})n\partial_t]\psi(x) = 0. \tag{135}$$

In Eq. (135), $\chi = |\boldsymbol{\chi}| = m_{\mathrm{eff}}$ plays the role of an effective rest mass of the photon field in the inhomogeneous medium. In our case $\partial_0 = n\partial_t$, and then let us assume that

$$\psi = \psi' \exp(-\mathrm{i}\, m_{\mathrm{eff}} t/n). \tag{136}$$

Obviously, under the conditions of $\chi_0 = \eta_0 = 0$, one has $\partial_t(m_{\mathrm{eff}}/n) = 0$. Moreover, let us take such a slowly-varying envelope approximations:

$$|\partial_t\psi'| \ll |\psi' m_{\mathrm{eff}}/n|, \quad |\partial_t^2\psi'| \ll |\psi' m_{\mathrm{eff}}/n|. \tag{137}$$

Substituting Eq. (136) into Eq. (135), and using Eq. (137) one has

$$\begin{aligned} \mathrm{i}n\frac{\partial}{\partial t}\psi' = -\frac{1}{2m_{\mathrm{eff}}}\nabla^2\psi' + \frac{1}{2m_{\mathrm{eff}}}[(\nabla\cdot\boldsymbol{\chi}) - \boldsymbol{\chi}\cdot\nabla - \Omega]\psi' \\ -\frac{1}{m_{\mathrm{eff}}}[\mathrm{i}\boldsymbol{\Sigma}\cdot(\boldsymbol{\chi}\times\nabla) + \mathrm{i}(\boldsymbol{\alpha}\cdot\boldsymbol{\chi})m_{\mathrm{eff}}]\psi' \end{aligned}. \quad (138)$$

The second line on the right-hand side of Eq. (138) is related to the contributions from the spin-orbit interactions. Eq. (138) can be regarded as optical analogues of the Dirac equation in the nonrelativistic limit. Here we have just presented a preliminary investigation on the spin-orbit interaction of photons. In our next work, based on this paper, we will present a thorough research on the spin-orbit interaction of photons.

**7. The Dirac-like equation in curved spacetime**

As we know, there is no unique vacuum sate in a general spacetime. For convenience, all of our discussions later will be presented from the point of view of a distant observer with respect to a black hole. People have shown that the free-space Maxwell equations in generally covariant form are equivalent to Maxwell's equations in a flat spacetime in the presence of an effective medium [64-68]. The $(1,0)\oplus(0,1)$ spinor description of the photon field allows us to study the effects of gravity on the photon field by means of spin connection and the tetrad formalism, which is of great advantage to study the gravitational spin-orbit coupling of the photon field. For example, without taking account of any quantum-mechanical effect, people have studied that the circular photon orbits in the Schwarzschild geometry [69]. In our formalism, however, quantum-mechanical corrections due to a gravitational spin-orbit coupling will be taken into account.

To present a self-contained material for all readers with different professional backgrounds, we will provide a systematical argument for introducing the general relativistic Dirac-like equation, based on spin connection and the tetrad formalism. In order to make our argument the more explicit as possible, let us first devote to an essential summary of the tetrad formalism. After some general considerations we restrict the further discussion to the special case of the Schwarzschild metric.

It is important to clarify the notations we are going to use throughout this section. We work in geometrized units, $\hbar = c = G = 1$, the metric signature is $(-,+,+,+)$. Differently from the previous chapters, small Latin indices $a$, $b$, . . . will range from 0 to 3, and will be used to denote tensor indices in the flat tangent space (namely, they are indices labeling tensor representations of the local Lorentz group, raised and lowered by the Minkowski metric). The Greek indices $\mu$, $\nu$, . . . also will range from 0 to 3, but they will refer to tensor objects defined on the Riemann manifold (hence they transform covariantly under diffeomorphisms, and are raised and lowered by the Riemann metric).

### 7.1 The tetrad formalism

The geometry of a Riemann space-time can always be locally approximated by the Minkowski geometry, such that we can always introduce, at any given point of a Riemann manifold, a "flat" tangent manifold described by the Minkowski metric. Therefore, to locally characterize the geometry of a four-dimensional Riemann spacetime, we can introduce a tetrad which is a set of axes $\boldsymbol{\eta}_a = \{\boldsymbol{\eta}_0, \boldsymbol{\eta}_1, \boldsymbol{\eta}_2, \boldsymbol{\eta}_3\}$ attached to each point $x^\mu$ of spacetime. We will choose an orthonormal tetrad, where the axes form a locally inertial frame at each point, so that the dot products of the axes constitute the Minkowski metric $\eta_{ab} = \mathrm{diag}(-1,1,1,1)$ , i.e., $\boldsymbol{\eta}_a \cdot \boldsymbol{\eta}_b = \eta_{ab}$. In other words, the orthonormal tetrad forms an orthonormal basis in the local Minkowski space tangent to the Riemann spacetime at the given point $x^\mu$, and they are orthonormal with respect to the Minkowski metric $\eta_{ab}$ of the tangent manifold, i.e. they satisfy the condition $\boldsymbol{\eta}_a \cdot \boldsymbol{\eta}_b = \eta_{ab}$.

The vierbein $V_a{}^\mu$ is defined to be the matrix that transforms between the tetrad frame and the coordinate frame (the tetrad index $a$ comes first, then the coordinate index $\mu$): $\boldsymbol{\eta}_a = V_a{}^\mu \boldsymbol{g}_\mu$, where the basis of coordinate tangent vectors $\boldsymbol{g}_\mu = \{\boldsymbol{g}_0, \boldsymbol{g}_1, \boldsymbol{g}_2, \boldsymbol{g}_3\}$ satisfy $\boldsymbol{g}_\mu \cdot \boldsymbol{g}_\nu = g_{\mu\nu}$. The inverse vierbein $V^a{}_\mu$ is defined to be the matrix inverse of the vierbein $V_a{}^\mu$, so that $V^a{}_\mu V_a{}^\nu = \delta^\nu_\mu$, $V^b{}_\mu V_a{}^\mu = \delta^b_a$. Then one has $\boldsymbol{g}_\mu = V^a{}_\mu \boldsymbol{\eta}_a$. The scalar spacetime distance is $\mathrm{d}s^2 = g_{\mu\nu}\mathrm{d}x^\mu \mathrm{d}x^\nu = \boldsymbol{g}_\mu \cdot \boldsymbol{g}_\nu \mathrm{d}x^\mu \mathrm{d}x^\nu = \eta_{ab} V^a{}_\mu V^b{}_\nu \mathrm{d}x^\mu \mathrm{d}x^\nu$, from which it follows that the coordinate metric $g_{\mu\nu}$ is $g_{\mu\nu} = \eta_{ab} V^a{}_\mu V^b{}_\nu$.

A tetrad transformation is a Lorentz transformation which may be a different transformation at each point. Then the tetrads $\boldsymbol{\eta}_a$ refer to axes that transform under local Lorentz transformations. As compared with which, the coordinate tangent vectors $\boldsymbol{g}_\mu$ refer to axes that transform under general coordinate transformations. Just as coordinate vectors (and tensors) were defined as objects that transformed like (tensor products of) coordinate intervals under coordinate transformations, so also tetrad vectors (and tensors) are defined as objects that transform like (tensor products of) tetrad vectors under tetrad (Lorentz) transformations. Likewise, just as the indices on a coordinate vector or tensor were lowered and raised with the coordinate metric $g_{\mu\nu}$ and its inverse $g^{\mu\nu}$, so also indices on a tetrad vector or tensor are lowered and raised with the tetrad metric $\eta_{ab}$ and its inverse $\eta^{ab}$.

Tetrad transformations rotate the tetrad axes $\boldsymbol{\eta}_a$ at each point by a Lorentz transformation $\Lambda_a{}^b$, while keeping the background coordinates $x^\mu$ unchanged:

$$\boldsymbol{\eta}_a \to \boldsymbol{\eta}'_a = \Lambda_a{}^b \boldsymbol{\eta}_b . \tag{139}$$

Lorentz transformations are precisely those transformations that leave the tetrad metric unchanged:

$$\eta_{ab} = \boldsymbol{\eta}_a \cdot \boldsymbol{\eta}_b \to \eta'_{ab} = \boldsymbol{\eta}'_a \cdot \boldsymbol{\eta}'_b = \Lambda_a{}^c \Lambda_b{}^d \boldsymbol{\eta}_c \cdot \boldsymbol{\eta}_d = \Lambda_a{}^c \Lambda_b{}^d \eta_{cd} = \eta_{ab} . \qquad (140)$$

A 4-vector can be written in a coordinate- and tetrad- independent fashion as an abstract 4-vector $\boldsymbol{A} = \boldsymbol{\eta}_a A^a = \boldsymbol{g}_\mu A^\mu$, its tetrad and coordinate components are related by the vierbein: $A_a = V_a{}^\mu A_\mu$, $A_\mu = V^a{}_\mu A_a$. The scalar product of two 4-vectors may be $\boldsymbol{A}$ and $\boldsymbol{B}$ may be written variously: $\boldsymbol{A} \cdot \boldsymbol{B} = A_a B^a = A_\mu B^\mu$ .

A coordinate vector $A^\mu$ (with a Greek index called curved index) does not change under a tetrad transformation (i.e., a local Lorentz (tangent-space) transformation), and is therefore a tetrad scalar; A tetrad vector $A^a$ (with a Latin index called flat index) does not change under a general coordinate transformation, and is therefore a coordinate scalar. An abstract vector $\boldsymbol{A}$, identified by boldface, is the thing itself, and is unchanged by either the choice of coordinates or the choice of tetrad. Since the abstract vector is unchanged by either a coordinate transformation or a tetrad transformation, it is a coordinate and tetrad scalar, and has no indices. Switch between coordinate and tetrad frames with the vierbein $V_a{}^\mu$ and its inverse $V^a{}_\mu$. The vierbeins transform as general-covariant vectors with respect to their curved index, and as Lorentz vectors with respect to the flat index:

$$V_a{}^\mu \to V_a{}'^\mu = \frac{\partial x'^\mu}{\partial x^\nu} \Lambda_a{}^b V_b{}^\nu . \qquad (141)$$

Directed derivatives $\boldsymbol{e}_a$ are defined to be the directional derivatives along the axes $\boldsymbol{\eta}_a$:

$$\boldsymbol{e}_a = \boldsymbol{\eta}_a \cdot \partial = \boldsymbol{\eta}_a \cdot \boldsymbol{g}^\mu \partial_\mu = V_a{}^\mu \partial_\mu . \qquad (142)$$

Obviously, $\boldsymbol{e}_a$ is a tetrad 4-vector. Unlike coordinate derivatives $\partial_\mu = \partial / \partial x^\mu$, directed derivatives $\boldsymbol{e}_a$ do not commute, their commutator is

$$[\boldsymbol{e}_a, \boldsymbol{e}_b] = [V_a{}^\mu \partial_\mu, V_b{}^\nu \partial_\nu] = [V_a{}^\nu (\partial_\nu V_b{}^\mu) - V_b{}^\nu (\partial_\nu V_a{}^\mu)]\partial_\mu \equiv C_{ab}{}^c \boldsymbol{e}_c . \qquad (143)$$

Let $C_{abc} = \eta_{dc} C_{ab}{}^d$ , using Eq. (143), $\boldsymbol{e}_c = V_c{}^\mu \partial_\mu$ and $V^b{}_\mu V_a{}^\mu = \delta^b_a$, one has:

$$C_{abc} = \eta_{dc} C_{ab}{}^d = \eta_{dc} V^d{}_\mu [V_a{}^\nu (\partial_\nu V_b{}^\mu) - V_b{}^\nu (\partial_\nu V_a{}^\mu)] . \qquad (144)$$

In a word, the equivalence principle requires local spacetime structure be identified with Minkowski spacetime possessing Lorentz symmetry. In order to relate local Lorentz symmetry to curved spacetime, one need to solder the local (tangent) space to the external (curved) space. The soldering tools are tetrad fields, and one can use the tetrad formalism to treat spinor fields in general relativity. To do so, one can construct a vierbein field $V_a{}^\mu(x)$ at every point in spacetime, it is a set of four orthonormal vectors which defines a frame, and because of the equivalence principle this frame can be made inertial at every point. Physical quantities have separate transformation properties in world space (with metric $g_{\mu\nu}$) and tangent space (with metric $\eta_{ab}$). The vierbein field $V_a{}^\mu(x)$ is a contravariant vector in world space and a covariant vector in tangent space.

### 7.2 Tetrad covariant derivative

The geometric description of gravity has been developed using the notions of Riemannian metric $g$ and Christoffel connection $\Gamma$, where the curvature of the spacetime manifold, its dynamical evolution, and its interaction with the matter sources has been described in terms of differential equations for the variables $g$ and $\Gamma$ . On the other hand, there has been an alternative (but fully equivalent) approach to the description of a Riemannian manifold based on the notions of vierbein $V$ and Lorentz connection (also called spin connection) $\Omega$ [70]. This alternative language is particularly appropriate to embed spinor fields in a curved spacetime. This alternative geometric formalism naturally leads to the formulation of general relativity as a gauge theory for a local symmetry group, thus putting gravity on the same footing of the other fundamental interactions. In particular, the gauge symmetry of the gravitational interactions is the local Lorentz symmetry, and the spacetime curvature can be interpreted as the Yang–Mills field for the Lorentz connection, which plays the role of the (non-Abelian) gauge potential.

By means of the projection transforming curved into flat indices, we can move from the diffeomorphisms of the Riemann manifold to the Lorentz transformations on the tangent manifold. The tangent manifold, however, varies from point to point, so that the corresponding Lorentz transformations are local transformations. The requirement of general covariance thus translates into the requirement of local Lorentz invariance.

A general-covariant geometric model, adapted to a curved space–time manifold, must be locally Lorentz invariant if referred to the tangent-space manifold described by the vierbein formalism. The proper, orthochronous Lorentz group is a 6-parameter Lie group, and a generic transformation can be represented in exponential form as follows:

$$L(\Lambda) = \exp(-\mathrm{i}\, \omega_{ab} S^{ab} / 2) . \tag{145}$$

The matrix $\omega_{ab} = -\omega_{ba}$ is antisymmetric and contains six real parameters, while the six generators $S_{ab} = -S_{ba}$ satisfy the Lie algebra of SO(3, 1):

$$\mathrm{i}[S_{ab}, S_{cd}] = \eta_{bc} S_{ad} - \eta_{ac} S_{bd} + \eta_{db} S_{ca} - \eta_{da} S_{cb} . \tag{146}$$

In order to restore the symmetry for local transformations with $\omega_{ab} = \omega_{ab}(x)$, we must associate the six generators $S_{ab}$ with six independent gauge vectors, i.e., the Lorentz connection (or spin connection) $\Omega_\mu$,

$$\Omega_\mu = \Omega^{ab}{}_\mu S_{ab} / 2 = \Omega_{ab\mu} S^{ab} / 2 , \quad \Omega^{ab}{}_\mu = -\Omega^{ba}{}_\mu , \tag{147}$$

and introduce a Lorentz covariant derivative defined by:

$$D_\mu = \partial_\mu - \mathrm{i}\Omega_\mu = \partial_\mu - \mathrm{i}\Omega^{ab}{}_\mu S_{ab}/2 \quad , \tag{148}$$

The tetrad-frame formulae look entirely similar to the coordinate-frame formulae, with the replacement of coordinate partial derivatives by directed derivatives, $\partial_\mu = \partial/\partial x^\mu \to \boldsymbol{e}_a$, and the replacement of coordinate-frame connections by tetrad-frame connections. By means of the vierbein $V_a{}^\mu$ and its inverse $V^a{}_\mu$ one can switch between coordinate and tetrad frames. Using Eqs. (142) and (148), one can show that the tetrad-frame spin connection is

$$D_a = V_a{}^\mu D_\mu = \boldsymbol{e}_a - \mathrm{i}\Gamma_{bca} S^{bc}/2 , \tag{149}$$

where $\Gamma_{bca} = V_a{}^\mu \Omega_{bc\mu}$. Both $\Gamma_{bca}$ and $\Omega^{ab}{}_\mu$ are called the connection coefficients. In terms of $C_{abc} = \eta_{dc} C_{ab}{}^d$ defined by Eqs. (143) and (144), one can express the connection coefficients $\Gamma_{bca}$ as

$$\Gamma_{bca} = -(C_{bca} + C_{cab} - C_{abc})/2 . \tag{150}$$

Using Eqs. (143) and (144) one can calculate $\Gamma_{bca}$ via Eq. (150).

**7.3 Dirac-like equation in Schwarzschild metric**

As for the photon field in the $(1,0)\oplus(0,1)$ spinor representation, the infinitesimal generators satisfying the Lie algebra given by Eq. (146) are

$$S_{lm} = \varepsilon_{lmn}\Sigma^n , \quad S_{0l} = -\mathrm{i}\alpha_l , \quad l,m,n = 1,2,3 , \tag{151}$$

where the matrices $\Sigma_l$ and $\alpha_l$ ($l = 1,2,3$) are given by Eq. (40) or Eq. (41). The Dirac-like equation in flat Minkowski spacetime is invariant under arbitrary global Lorentz transformations. Within the framework of general relativity, the Dirac-like equation in flat Minkowski spacetime should be replaced by the one in a curved spacetime, such that the resulting equation is invariant under local Lorentz transformations. According to the discussions above, using the covariant derivative $D_a$ given by Eq. (149), one can express the Dirac-like equation in a curved spacetime as ($\hbar = c = G = 1$):

$$\mathrm{i}\beta^a D_a \psi(x) = \mathrm{i}\beta^a (\boldsymbol{e}_a - \mathrm{i}\Gamma_{bca} S^{bc}/2)\psi(x) = 0\,, \tag{152}$$

where $\beta^a$'s are given by Eq. (40) or Eq. (41), and $S^{ab}$ are given by Eq. (151). For the moment, the photon field $\psi(x)$ transforms like a scalar with respect to 'world' transformations, but a $(1,0)\oplus(0,1)$ spinor with respect to Lorentz transformations in tangent space.

From now on we will focus on the special case of the Schwarzschild metric. Outside a Schwarzschild black hole of mass *M*, the standard form of the Schwarzschild metric is ($\hbar = c = G = 1$)

$$\mathrm{d}s^2 = -(1 - r_s/r)\mathrm{d}t^2 + (1 - r_s/r)^{-1}\mathrm{d}r^2 + r^2(\mathrm{d}\theta^2 + \sin^2\theta \mathrm{d}\phi^2)\,, \tag{153}$$

where $r_s = 2M$ is the Schwarzschild radius of the black hole. One can introduce a new radial coordinate [71]

$$\rho = (r - r_s/2 + \sqrt{r^2 - r_s r})/2\,,\ \text{ or }\ r = \rho(1 + r_s/4\rho)^2\,, \tag{154}$$

which implies that $\rho \to r_s/4$ for $r \to r_s$, and the Schwarzschild metric reads:

$$\mathrm{d}s^2 = -(1 - r_s/4\rho)^2(1 + r_s/4\rho)^{-2}c^2\mathrm{d}t^2 + (1 + r_s/4\rho)^4(\mathrm{d}\rho_1^2 + \mathrm{d}\rho_2^2 + \mathrm{d}\rho_3^2)\,, \tag{155}$$

where the variables $\rho_1$, $\rho_2$ and $\rho_3$ are defined by $\rho_1 = \rho\sin\theta\cos\phi$, $\rho_2 = \rho\sin\theta\sin\phi$, $\rho_3 = \rho\cos\theta$, respectively. Eqs. (153) and (155) can be written in a unified form:

$$\mathrm{d}s^2 = -a_0^2\mathrm{d}t^2 + a_1^2\mathrm{d}x_1^2 + a_2^2\mathrm{d}x_2^2 + a_3^2\mathrm{d}x_3^2\,. \tag{156}$$

For Eq. (153) one has $x^\mu = (t, \boldsymbol{x})$ with $x^0 = t$, $x^1 = r$, $x^2 = \theta$ and $x^3 = \phi$, and

$$a_0 = (1 - r_s/r)^{1/2}\,,\ \ a_1 = (1 - r_s/r)^{-1/2}\,,\ \ a_2 = r\,,\ \ a_3 = r\sin\theta\,. \tag{157}$$

For Eq. (155) one has $x^\mu = (t, \boldsymbol{x})$ with $x^0 = t$, $\boldsymbol{x} = \boldsymbol{\rho} = (\rho_1, \rho_2, \rho_3)$, $\rho = |\boldsymbol{\rho}|$, and

$$a_0 = (1 - r_s/4\rho)(1 + r_s/4\rho)^{-1}, \quad a = a_1 = a_2 = a_3 = (1 + r_s/4\rho)^2. \tag{158}$$

Eq. (155) is the so-called isotropic form of the metric in Schwarzschild spacetime, the gradient operator $\nabla = (\partial_1, \partial_2, \partial_3)$ based on Eq. (155) satisfies $\partial_l = \partial/\partial\rho^l$, $l = 1, 2, 3$.

To apply the connection coefficients in an orthonormal basis, let us rewrite Eq. (156) as

$$\mathrm{d}s^2 = -(\boldsymbol{\theta}^0)^2 + (\boldsymbol{\theta}^1)^2 + (\boldsymbol{\theta}^2)^2 + (\boldsymbol{\theta}^3)^2, \tag{159}$$

where $\boldsymbol{\theta}^0 = a_0\mathbf{d}t$, $\boldsymbol{\theta}^l = a_l\mathbf{d}x^l$ ( $l = 1, 2, 3$ ) with $g_{\mu\nu} = \eta_{\mu\nu} = \mathrm{diag}(-1,1,1,1)$ form an orthonormal basis, For the moment, we do not distinguish between tensor indices in the flat tangent space and those in Riemannian spacetime, and denote them by the Greek indices uniformly. The dual basis of $\boldsymbol{\theta}^\mu = a_\mu\mathbf{d}x^\mu$ is

$$\boldsymbol{e}_\mu = a_\mu^{-1}\,\partial/\partial x^\mu = a_\mu^{-1}\partial_\mu, \quad \mu = 0, 1, 2, 3. \tag{160}$$

In the dual vector space Eqs. (143) and (144) become:

$$[\boldsymbol{e}_\kappa, \boldsymbol{e}_\lambda] = C_{\kappa\lambda}{}^{\mu}\boldsymbol{e}_\mu, \quad C_{\kappa\lambda\nu} = \eta_{\mu\nu}C_{\kappa\lambda}{}^{\mu}, \tag{161}$$

Likewise, in the orthonormal basis with with $g_{\mu\nu} = \eta_{\mu\nu} = \mathrm{diag}(-1,1,1,1)$, the connection coefficients given by Eq. (150) can be rewritten as

$$\Gamma_{\kappa\lambda\mu} = -(C_{\kappa\lambda\mu} + C_{\lambda\mu\kappa} - C_{\mu\kappa\lambda})/2. \tag{162}$$

In the Schwarzschild spacetime, the Dirac-like equation Eq. (152) can be rewritten as

$$\mathrm{i}\beta^\mu(\boldsymbol{e}_\mu - \mathrm{i}\Gamma_{\kappa\lambda\mu}S^{\kappa\lambda}/2)\psi(x) = 0. \tag{163}$$

Using Eqs. (156), (160) and (161), one can prove that ( $l, m = 1, 2, 3$ and $l \neq m$ )

$$C_{0l0} = -C_{l00} = -a_l^{-1}\partial_l \ln a_0, \quad C_{lmm} = -C_{mlm} = -a_l^{-1}\partial_l \ln a_m, \tag{164}$$

with the others vanishing. Using Eqs. (162) and (164), seeing that $S^{\mu\nu} = -S^{\nu\mu}$ and $C_{\kappa\lambda\nu} = -C_{\lambda\kappa\nu}$, one can prove that

$$\begin{aligned}\beta^\mu\Gamma_{\kappa\lambda\mu}S^{\kappa\lambda}/2 = &-\beta^0 S^{l0}a_l^{-1}\partial_l \ln a_0 + \beta^2 S^{12}a_1^{-1}\partial_1 \ln a_2 + \beta^3 S^{13}a_1^{-1}\partial_1 \ln a_3 \\ &+ \beta^1 S^{21}a_2^{-1}\partial_2 \ln a_1 + \beta^3 S^{23}a_2^{-1}\partial_2 \ln a_3 + \beta^1 S^{31}a_3^{-1}\partial_3 \ln a_1 + \beta^2 S^{32}a_3^{-1}\partial_3 \ln a_2\end{aligned}. \tag{165}$$

Substituting Eqs. (160) and (165) into Eq. (163), one can obtain

$$\begin{aligned}&a_0^{-1}\mathrm{i}\beta^0\partial_0\psi+a_1^{-1}[\mathrm{i}\beta^1\partial_1-\beta^0S^{10}\partial_1\ln a_0+\beta^2S^{12}\partial_1\ln a_2-\beta^3S^{31}\partial_1\ln a_3]\psi\\&+a_2^{-1}[\mathrm{i}\beta^2\partial_2-\beta^0S^{20}\partial_2\ln a_0+\beta^3S^{23}\partial_2\ln a_3-\beta^1S^{12}\partial_2\ln a_1]\psi\\&+a_3^{-1}[\mathrm{i}\beta^3\partial_3-\beta^0S^{30}\partial_3\ln a_0+\beta^1S^{31}\partial_3\ln a_1-\beta^2S^{23}\partial_3\ln a_2]\psi=0\end{aligned}\quad. \tag{166}$$

Using Eq. (40) or (41), $[\tau_l,\tau_m]=\mathrm{i}\varepsilon_{lmn}\tau_n$, one can prove that

$$\beta^l\Sigma^m-\beta^m\Sigma^l=\mathrm{i}\varepsilon^{lmn}\beta_n,\quad \beta^l\Sigma^m-\Sigma^m\beta^l=\mathrm{i}\varepsilon^{lmn}\beta_n. \tag{167}$$

Using Eqs. (151) and (167), $\beta^0\alpha^l=\beta^l$, one can show that Eq. (166) becomes

$$\begin{aligned}&a_0^{-1}\mathrm{i}\beta^0\partial_0\psi+a_1^{-1}\mathrm{i}\beta^1[\partial_1+(\partial_1\ln a_0a_2)]\psi+a_2^{-1}\mathrm{i}\beta^2[\partial_2+(\partial_2\ln a_0a_3)]\psi\\&+a_3^{-1}\mathrm{i}\beta^3[\partial_3+(\partial_3\ln a_0a_1)]\psi+a_1^{-1}\beta^3\Sigma^2[\partial_1\ln(a_2/a_3)]\psi\\&+a_2^{-1}\beta^1\Sigma^3[\partial_2\ln(a_3/a_1)]\psi+a_3^{-1}\beta^2\Sigma^1[\partial_3\ln(a_1/a_2)]\psi=0\end{aligned}\quad. \tag{168}$$

Let us discuss Eq. (168) as follows:

1) Substituting Eq. (157) into Eq. (168) and using Eq. (167), one has

$$\begin{aligned}\mathrm{i}\frac{\partial}{\partial t}\psi=&-\mathrm{i}\alpha^1(1-\frac{r_s}{r})(\frac{\partial}{\partial r}+\frac{1}{r})\psi-\mathrm{i}\alpha^2(1-\frac{r_s}{r})^{1/2}\frac{1}{r}(\frac{\partial}{\partial\theta}+\cot\theta)\psi\\&-\mathrm{i}\alpha^3(1-\frac{r_s}{r})^{1/2}\frac{1}{r\sin\theta}\frac{\partial}{\partial\phi}\psi-\mathrm{i}\alpha^1\frac{r_s}{2r^2}\psi-\alpha^1\Sigma^3(1-\frac{r_s}{r})^{1/2}\frac{\cot\theta}{r}\psi\end{aligned}\quad. \tag{169}$$

2) Substituting Eq. (158) into Eq. (168) and using Eq. (167), one has

$$\mathrm{i}\beta^0\eta\partial_0\psi+\mathrm{i}\beta^l(\partial_l+\Pi_l)\psi(x)=0, \tag{170}$$

where

$$\eta=a_0^{-1}a=(1-r_s/4\rho)^{-1}(1+r_s/4\rho)^3, \tag{171}$$

$$\boldsymbol{\Pi}=\nabla\ln\sqrt{\varpi}=\nabla\ln(1-r_s/4\rho)(1+r_s/4\rho), \tag{172}$$

$$\varpi=(a_0a)^2=(1-r_s/4\rho)^2(1+r_s/4\rho)^2. \tag{173}$$

Let $\psi=\varpi^{-1/2}\varphi$, $\partial_0'=\eta\,\partial/\partial t$, $\partial_\mu'=(\partial_0',\nabla)=(\eta\partial_0,\partial_l)$, Eq. (170) can be rewritten as

$$\mathrm{i}\beta^\mu\partial_\mu'\varphi=0. \tag{174}$$

Note that for $\psi=\psi_{\mathrm{S}}$, $\psi_{\mathrm{C}}$, one has $\psi_{\mathrm{S}}=\varpi^{-1/2}\varphi_{\mathrm{S}}$ and $\psi_{\mathrm{C}}=\varpi^{-1/2}\varphi_{\mathrm{C}}$. According to the tetrad formalism, the metric tensor for Riemannian spacetime transforms like a scalar with respect to Lorentz transformations in tangent space. As a result, $\eta = a_0^{-1}a$ and $\varpi = (a_0 a)^2$ are two Lorentz scalars in the local Minkowski spacetime. To guarantee Eq. (170) be Lorentz covariant in the local Minkowski spacetime, the constraint conditions should be taken as

$$(\nabla + \boldsymbol{\Pi}) \cdot \boldsymbol{E} = 0, \quad (\nabla + \boldsymbol{\Pi}) \cdot \boldsymbol{H} = 0. \tag{175}$$

It follows from $\psi = \varpi^{-1/2}\varphi$ that $\varphi$ is formed by $\boldsymbol{E}' = \varpi^{1/2}\boldsymbol{E}$ and $\boldsymbol{H}' = \varpi^{1/2}\boldsymbol{H}$ in the same way as $\psi$ being formed by $\boldsymbol{E}$ and $\boldsymbol{H}$, and Eq. (175) becomes

$$\nabla \cdot \boldsymbol{E}' = \nabla \cdot \boldsymbol{H}' = 0. \tag{176}$$

In terms of $\varphi$ and $\bar{\varphi} = \varphi^{\dagger}\beta^{0}$, one can express Eq. (86) as

$$\Delta_{\mu\nu} = \bar{\varphi}[\mathrm{i}(\beta_{\nu}\partial'_{\mu} - \beta_{\mu}\partial'_{\nu}) + (\beta^{\rho}S_{\mu\nu} - S_{\mu\nu}\beta^{\rho})\partial'_{\rho}]\varphi = 0. \tag{177}$$

Because $\eta = a_0^{-1}a$ and $\varpi = (a_0 a)^2$ are two Lorentz scalars in tangent space (the local Minkowski spacetime), in the local Minkowski spacetime $\varphi$ is still the $(1,0)\oplus(0,1)$ spinor, such that the pseudo-Lagrangian density of $\mathcal{L} = \bar{\varphi}(x)(\mathrm{i}\beta^{\mu}\partial'_{\mu})\varphi(x)$ is a Lorentz scalar, and Eq. (174) is Lorentz covariant. Eq. (174) implies that, with the velocity of light in Minkowski vacuum replaced by $1/\eta$ in an equivalent medium (with the refractive index $\eta = a_0^{-1}a$), the Dirac-like equation in Schwarzschild spacetime is equivalent to the one in Minkowski spacetime.

Let

$$\boldsymbol{\Lambda} = \nabla \ln\sqrt{\eta} = \nabla \ln\sqrt{(1 - r_s/4\rho)^{-1}(1 + r_s/4\rho)^3}. \tag{178}$$

Using $\partial'_0 = \eta\,\partial/\partial t$, $\nabla\eta = 2\eta\boldsymbol{\Lambda}$, let $f = f(t, \boldsymbol{\rho})$ be a function, it is easy to show that

$$\nabla\partial'_0 f - \partial'_0\nabla f = 2\boldsymbol{\Lambda}\partial'_0 f, \quad \nabla \times \boldsymbol{\Lambda} f = -\boldsymbol{\Lambda} \times \nabla f, \quad \partial'_0\boldsymbol{\Lambda} f = \boldsymbol{\Lambda}\partial'_0 f. \tag{179}$$

$$\partial_0\boldsymbol{\Pi} f = \boldsymbol{\Pi}\partial_0 f, \quad \nabla \times \boldsymbol{\Pi} f = -\boldsymbol{\Pi} \times \nabla f. \tag{180}$$

Using Eq. (40) or (41), Eqs. (39) , (179) and (180), it follows from Eq. (171) that

$$\partial'^{0}\partial'_{0}\psi+(\nabla+\boldsymbol{\Pi})\cdot(\nabla+\boldsymbol{\Pi})\psi+2\boldsymbol{\alpha}\cdot\boldsymbol{\Lambda}\partial'_{0}\psi=0\,. \tag{181}$$

It is difficult to solve the exact solutions of Eq. (169), Eq. (170) or Eq. (174). However, for our purpose, we will just study a special case in next section.

### 7. 4 Gravitational spin-orbit interaction of the photon field

The existence of bound states within an atom induces people to consider a similar situation, namely the behavior of a particle within a spherical symmetric gravitational field (see for example, Ref. [72]). On the other hand, many investigations on the trajectories of light in the Schwarzschild metric have been presented [73, 74], from which people concluded that there are no stable circular photon orbits in the Schwarzschild geometry [69]. To present an application example of the $(1,0)\oplus(0,1)$ spinor description of the photon field, let us study the effect of gravitational spin-orbit coupling on the circular photon orbit in the Schwarzschild geometry.

To present a self-contained formalism, it is worth pausing to review the traditional investigations on the trajectory of a photon in the Schwarzschild metric Eq. (153). Let us define a 'Lagrangian'

$$\mathcal{L}=g_{\mu\nu}\dot{x}^{\mu}\dot{x}^{\nu}=g_{\mu\nu}\frac{\mathrm{d}x^{\mu}}{\mathrm{d}\sigma}\frac{\mathrm{d}x^{\nu}}{\mathrm{d}\sigma}, \tag{182}$$

where $\dot{x}^{\mu}\equiv\mathrm{d}x^{\mu}/\mathrm{d}\sigma$, and $\sigma$ is an affine parameter along the geodesic $x^{\mu}(\sigma)$. The geodesic equations can be obtained by substituting $\mathcal{L}$ into the Euler–Lagrange equations:

$$\frac{\mathrm{d}}{\mathrm{d}\sigma}\left(\frac{\partial\mathcal{L}}{\partial\dot{x}^{\mu}}\right)-\frac{\partial\mathcal{L}}{\partial x^{\mu}}=0\,. \tag{183}$$

For photon moving in the equatorial plane of the Schwarzschild geometry (and then $\theta=\pi/2$), using Eqs. (153), (182) and (183) one can obtain the 'energy' equation for photon orbits ($\hbar = c = G = 1$)

$$\dot{r}^2 + V_{\text{eff}}(r) = \omega^2 , \tag{184}$$

where the effective potential of $V_{\text{eff}}(r)$ is defined as (as compared with the definition of $V_{\text{eff}}(r)$ in Ref. [69], our definition has a additional multiplication factor $h^2$)

$$V_{\text{eff}}(r) = \frac{h^2}{r^2}(1 - \frac{r_s}{r}) . \tag{185}$$

In Eqs. (184) and (185) the quantities $\omega$ and $h$ are two constants, and they are defined as, respectively

$$\omega = (1 - \frac{r_s}{r})\frac{\mathrm{d}t}{\mathrm{d}\sigma} , \quad h = r^2 \frac{\mathrm{d}\phi}{\mathrm{d}\sigma} . \tag{186}$$

From the point of view of a distant observer (an observer at rest at infinity), $\omega$ is the total energy of a photon in its orbit, and $h$ equals the specific angular momentum of the photon. Let $u = 1/r$ , the analogue for photons of the shape equation can be obtained from Eqs. (184)-( 186):

$$\frac{\mathrm{d}^2 u}{\mathrm{d}\phi^2} + u = 3r_s u^2 / 2 . \tag{187}$$

For motion in a circle $r$ (and then $u = 1/r$ ) is a constant. Thus, Eq. (187) implies that the only possible radius for a circular photon orbit is $r = 3r_s/2 = 3M$ . For the moment, Eq. (184) becomes

$$\omega^2 = V_{\text{eff}}(r) = \frac{h^2}{r^2}(1 - \frac{r_s}{r}) = \frac{4h^2}{27 r_s^2} . \tag{188}$$

In Eq. (188) $h$ plays the role of the orbital angular momentum. However, one can show that $V_{\text{eff}}(r)$ has a single maximum at $r = 3r_s/2 = 3M$ , such that this circular photon orbit is not a stable one. In other words, there are no stable circular photon orbits in the Schwarzschild geometry.

To study the effect of a gravitational spin-orbit coupling on the circular photon orbits in the Schwarzschild geometry, let us first start from Eq. (169). Let $\psi=\xi\exp(-\mathrm{i}\omega t+\mathrm{i}m\phi)$, where $\omega>0$ and $m$ is an integer. For circular motion in the equatorial plane we have $\theta=\pi/2$ and $r$=constant, and so $\partial_r\xi=\partial_\theta\xi=0$, for the moment Eq. (169) becomes

$$\omega\xi=-\mathrm{i}\frac{1}{r}\alpha^1\xi+\mathrm{i}\frac{r_s}{2r^2}\alpha^1\xi+(1-\frac{r_s}{r})^{1/2}\frac{m}{r}\alpha^3\xi\,. \tag{189}$$

It is more convenient to solve Eq. (189) in the chiral representation. Let $\xi=\begin{pmatrix}f & g\end{pmatrix}^{\mathrm{T}}$ (T denotes the matrix transpose) and use Eq. (179), it follows from Eq. (189) that

$$\begin{cases}\omega f+\mathrm{i}r^{-1}\tau_1 f-\mathrm{i}\,r_s r^{-2}\tau_1 f/2-(1-r_s r^{-1})^{1/2}mr^{-1}\tau_3 f=0\\ \omega g-\mathrm{i}r^{-1}\tau_1 g+\mathrm{i}\,r_s r^{-2}\tau_1 g/2+(1-r_s r^{-1})^{1/2}mr^{-1}\tau_3 g=0\end{cases}. \tag{190}$$

Using Eq. (11) and the conditions of nonzero solution, it follows from Eq. (190) that

$$\omega^3-\omega(1-r_s r^{-1})m^2r^{-2}+\omega(r^{-1}-r_s r^{-2}/2)^2=0\,, \tag{191}$$

which implies that (where $\omega=0$, related to the longitudinal polarization solution, is discarded)

$$\omega^2=\frac{m^2}{r^2}(1-\frac{r_s}{r})-(\frac{2r-r_s}{2r^2})^2=\frac{m^2-1}{r^2}(1-\frac{r_s}{r})-\frac{r_s^2}{4r^4}\,. \tag{192}$$

Consider that $r>r_s$ and $\omega>0$, it follows from Eq. (192) that $m^2>1$ (i.e., $m^2\geq 4$). In the circular photon orbit with the radius of $r=3r_s/2$, Eq. (192) becomes

$$\omega^2=\frac{4m^2}{27r_s^2}-\frac{16}{81r_s^2}\,,\quad m^2\geq 4\,. \tag{193}$$

Consider that $m$ plays the role of the orbital angular momentum, comparing Eq. (192) (or Eq. (193)) with Eq. (188), one can interpret the second term on the right-hand side of Eq. (192) (or Eq. (193)) as the *average* contribution from the gravitational spin-orbit coupling. Note that $\xi$ in Eq. (189) cannot be the eigenstates of helicity, and then we obtain the average result of the gravitational spin-orbit coupling, which is different from the case we will discuss below.

Now, Let us study our issue based on Eq. (174). Taking the standard representation as an example, substituting $\sqrt{2}\varphi_{S} = \begin{pmatrix} \boldsymbol{E}' & \mathrm{i}\boldsymbol{H}' \end{pmatrix}^{\mathrm{T}}$ into Eq. (174) and using Eq. (178) one has

$$(\boldsymbol{\tau}\cdot\nabla)\boldsymbol{H}' = \mathrm{i}\eta\partial_t\boldsymbol{E}', \quad (\boldsymbol{\tau}\cdot\nabla)\boldsymbol{E}' = -\mathrm{i}\eta\partial_t\boldsymbol{H}'. \tag{194}$$

As for $\nabla = (\partial_1, \partial_2, \partial_3)$ with $\partial_l = \partial/\partial\rho_l$ ($l = 1,2,3$), its column-matrix form is denoted as $\boldsymbol{\lambda} = \begin{pmatrix} \partial_1 & \partial_2 & \partial_3 \end{pmatrix}^{\mathrm{T}}$, Eq. (176) implies that

$$\boldsymbol{\lambda}\boldsymbol{\lambda}^{\mathrm{T}}\boldsymbol{E}' = \boldsymbol{\lambda}\boldsymbol{\lambda}^{\mathrm{T}}\boldsymbol{H}' = 0. \tag{195}$$

Let Eq. (178) be rewritten as

$$\boldsymbol{\Lambda} = \nabla \ln\sqrt{(1 - r_s/4\rho)^{-1}(1 + r_s/4\rho)^3} = \boldsymbol{e}_{\rho}\Lambda_{\rho}, \tag{196}$$

where $\boldsymbol{e}_{\rho} = \boldsymbol{\rho}/\rho$, and

$$\Lambda_{\rho} = -\frac{r_s}{8\rho^2}\left[\frac{3}{(1 + r_s/4\rho)} + \frac{1}{(1 - r_s/4\rho)}\right]. \tag{197}$$

Let $f = \boldsymbol{E}', \boldsymbol{H}'$, using Eqs. (11), (195) and $\nabla\eta = 2\eta\boldsymbol{\Lambda}$, one can obtain from Eq. (194) that

$$\eta^2\partial_t^2 f = \nabla^2 f - 2\boldsymbol{\Lambda}\cdot\nabla f - 2\mathrm{i}\boldsymbol{\tau}\cdot(\boldsymbol{\Lambda}\times\nabla) f. \tag{198}$$

The last term on the right-hand side of Eq. (198) represents the spin-orbit coupling interaction. Let $f = F\exp(-\mathrm{i}\omega t)$, where $\omega > 0$, Eq. (198) gives

$$\nabla^2 F - 2\boldsymbol{\Lambda}\cdot\nabla F - 2\mathrm{i}\boldsymbol{\tau}\cdot(\boldsymbol{\Lambda}\times\nabla)F + \eta^2\omega^2 F = 0, \tag{199}$$

Seeing that there is a spherical symmetry and $\rho_1 = \rho\sin\theta\cos\phi$, $\rho_2 = \rho\sin\theta\sin\phi$, $\rho_3 = \rho\cos\theta$, let

$$\begin{cases} \tau_{\rho} = \tau_1\sin\theta\cos\phi + \tau_2\sin\theta\sin\phi + \tau_3\cos\theta \\ \tau_{\theta} = \tau_1\cos\theta\cos\phi + \tau_2\cos\theta\sin\phi - \tau_3\sin\theta \\ \tau_{\phi} = -\tau_1\sin\phi + \tau_2\cos\phi \end{cases}. \tag{200}$$

For the moment Eq. (199) can be rewritten as

$$\begin{aligned}&\frac{1}{\rho^2}\frac{\partial}{\partial\rho}(\rho^2\frac{\partial}{\partial\rho}F)+\frac{1}{\rho^2\sin\theta}\frac{\partial}{\partial\theta}(\sin\theta\frac{\partial}{\partial\theta}F)+\frac{1}{\rho^2\sin^2\theta}\frac{\partial^2}{\partial\phi^2}F\\&+\eta^2\omega^2F-2\Lambda_\rho\frac{\partial F}{\partial\rho}-2\mathrm{i}\Lambda_\rho(\tau_\phi\frac{1}{\rho}\frac{\partial F}{\partial\theta}-\tau_\theta\frac{1}{\rho\sin\theta}\frac{\partial F}{\partial\phi})=0\end{aligned}. \tag{201}$$

Let (let $\hbar$ appear temporarily)

$$\hat{p}_\rho=-\mathrm{i}\hbar(\frac{\partial}{\partial\rho}+\frac{1}{\rho}),\quad \hat{\boldsymbol{L}}^2=-\hbar^2[\frac{1}{\sin\theta}\frac{\partial}{\partial\theta}(\sin\theta\frac{\partial}{\partial\theta})+\frac{1}{\sin^2\theta}\frac{\partial^2}{\partial\phi^2}]. \tag{202}$$

It follows from Eq. (202) that

$$[\rho,\hat{p}_\rho]=\mathrm{i}\hbar,\quad \hat{p}_\rho^2=-\hbar^2(\frac{\partial^2}{\partial\rho^2}+\frac{2}{\rho}\frac{\partial}{\partial\rho})=-\hbar^2\frac{1}{\rho^2}\frac{\partial}{\partial\rho}(\rho^2\frac{\partial}{\partial\rho}). \tag{203}$$

One can express Eq. (201) in terms of $\hat{p}_r^2$ and $\hat{\boldsymbol{L}}^2$:

$$-\frac{\hat{p}_\rho^2}{\hbar^2}F-\frac{\hat{\boldsymbol{L}}^2}{\hbar^2\rho^2}F+\eta^2\omega^2F-2\Lambda_\rho\frac{\partial F}{\partial\rho}-2\mathrm{i}\Lambda_\rho\frac{1}{\rho}(\tau_\phi\frac{\partial F}{\partial\theta}-\tau_\theta\frac{1}{\sin\theta}\frac{\partial F}{\partial\phi})=0. \tag{204}$$

Let $F=\zeta\exp(\mathrm{i}m\phi)$, where $m$ is actually an integer. For circular motion in the equatorial plane one has $\theta=\pi/2$ and $\rho$=constant (and then $r$=constant), and so $\partial_\rho\zeta=\partial_\theta\zeta=0$, $\tau_\theta=-\tau_3$ (see Eq. (200)), for the moment Eq. (201) (or Eq. (204)) becomes

$$(2m\rho^{-1}\Lambda_\rho\tau_3+\eta^2\omega^2-m^2\rho^{-2})\zeta=0. \tag{205}$$

The nonzero-solution conditions of Eq. (205) imply that (where $\eta^2\omega^2=m^2\rho^{-2}$, related to the longitudinal polarization solution, is discarded)

$$\omega^2=\frac{m^2}{\eta^2\rho^2}\pm\frac{2m}{\eta^2\rho}\Lambda_\rho. \tag{206}$$

The second term on the right-hand side of Eq. (206) comes from the contribution of the spin-orbit coupling interaction. In the circular photon orbit with the radius of $r=3r_s/2$, using Eqs. (154), (171) and (197), one has

$$\rho=(2+\sqrt{3})r_s/4,\quad \eta=12\sqrt{3}-18,\quad \Lambda_\rho=-2(2-\sqrt{3})/r_s. \tag{207}$$

Because of $\omega>0$, it follows from Eq. (206) that $m^2\geq 4$. Substituting Eq. (207) into Eq.

(204), one has

$$\omega^2 \equiv \omega_\pm^2 = \frac{4}{27r_s^2} m(m \pm 1) = \frac{4m^2}{27r_s^2} \pm \frac{4m}{27r_s^2}, \tag{208}$$

where $\omega_+ = 4m(m+1)/27r_s^2$ and $\omega_- = 4m(m-1)/27r_s^2$. Note that both $h$ and $m$ play the roles of the orbital angular momentum, comparing Eq. (208) with Eq. (188), one can show that the second term on the right-hand side of Eq. (208) comes from the contribution of the gravitational spin-orbit coupling interaction. To be specific, $\omega_+$ and $\omega_-$ represent the energies of photons with the spin projections of $\pm 1$, respectively. Therefore, the spin-orbit coupling interaction induces a splitting of energy levels.

More specifically, once the gravitational spin-orbit coupling is taken into account, the radius of the circular photon orbit is no longer $r = 3r_s/2 = 3M$. To present a heuristic discussion for this, let us present an equivalent description for the process from Eq. (184) to Eq. (188). Eq. (184) is rewritten as

$$\omega^2 = (\frac{\mathrm{d}r}{\mathrm{d}\sigma})^2 + \frac{h^2}{r^2}(1 - \frac{r_s}{r}). \tag{209}$$

For motion in a circle $r$ is a constant, Eq. (209) becomes

$$\omega^2 = V_{\mathrm{eff}}(r) = \frac{h^2}{r^2}(1 - \frac{r_s}{r}). \tag{210}$$

Let $\mathrm{d}V_{\mathrm{eff}}(r)/\mathrm{d}r = 0$, using Eq. (210) one has

$$r = 3r_s/2 = 3M. \tag{211}$$

Eq. (211) gives the radius of the circular photon orbit.

Eqs. (192) and (206) can be regarded as the quantum-mechanical (semi-classical) counterparts of Eq. (210) with the correspondence of $m \leftrightarrow h$ ($h^2 > 1$). As a result, when we consider the average effect of gravitational spin-orbit coupling, we rewrite Eq. (192) as,

$$\omega^2(r) = \frac{h^2}{r^2}(1 - \frac{r_s}{r}) - (\frac{2r - r_s}{2r^2})^2 = \frac{h^2 - 1}{r^2}(1 - \frac{r_s}{r}) - \frac{r_s^2}{4r^4}. \tag{212}$$

Let $\mathrm{d}\omega^2/\mathrm{d}r = 0$, using Eq. (212) and consider that $r > r_s$, one has

$$r = \frac{3r_s}{4} + \frac{r_s\sqrt{9(h^2-1)^2 + 8(h^2-1)}}{4(h^2-1)}. \tag{213}$$

Because of the gravitational spin-orbit coupling, the radius is related to the orbital angular momentum *h*. One can prove that $\mathrm{d}^2\omega^2/\mathrm{d}r^2 < 0$, that is, $\omega^2$ has a single maximum at the radius given by Eq. (213).

Likewise, when we consider the gravitational spin-orbit coupling interaction of photons with the helicities of $\pm 1$, respectively, using Eqs. (171) and (197) we rewrite Eq. (206) as

$$\omega_+^2(\rho) = \frac{h(1 - r_s/4\rho)^2}{\rho^2(1 + r_s/4\rho)^6}[h - (\frac{3r_s}{4\rho + r_s} + \frac{r_s}{4\rho - r_s})]. \tag{214}$$

$$\omega_-^2(\rho) = \frac{h(1 - r_s/4\rho)^2}{\rho^2(1 + r_s/4\rho)^6}[h + (\frac{3r_s}{4\rho + r_s} + \frac{r_s}{4\rho - r_s})]. \tag{215}$$

Starting from $\mathrm{d}\omega_\pm^2/\mathrm{d}\rho = 0$ one can obtain two quartic equations about $\rho > r_s/4$, each of them gives only one meaningful solution. As an example, let $r_s = 1$, $h = 2$, Eq. (207) implies that, without taking into account the gravitational spin-orbit coupling interaction,

$$\rho_0 = (2 + \sqrt{3})/4 \approx 0.933. \tag{216}$$

On the other hand, substituting $r_s = 1$ and $h = 2$ into $\mathrm{d}\omega_+^2/\mathrm{d}\rho = 0$, one can obtain that

$$64\rho^3 - 112\rho^2 + 40\rho - 3 = 0 \Rightarrow \rho_+ \approx 1.295 > \rho_0 \approx 0.933. \tag{217}$$

Likewise, substituting $r_s = 1$ and $h = 2$ into $\mathrm{d}\omega_+^2/\mathrm{d}\rho = 0$, one can obtain that

$$128\rho^4 - 32\rho^3 - 80\rho^2 + 22\rho - 1 = 0 \Rightarrow \rho_- \approx 0.783 < \rho_0 \approx 0.933. \tag{218}$$

Therefore, when the gravitational spin-orbit coupling is ignored, the radius of the circular photon orbit satisfies $\rho_0 = (2 + \sqrt{3})r_s/4$ (corresponding to $r = 3r_s/2$); once the gravitational spin-orbit coupling is taken into account, the radiuses of circular photon orbits satisfy $\rho_+ > \rho_0$ and $\rho_- < \rho_0$, for photons with the helicities of $\pm 1$, respectively, which is also related to the fact that the spin-orbit coupling interaction induces a splitting of energy levels. Similar to the effective potential $V_{\mathrm{eff}}(r)$ given by Eq. (210), $\omega_{\pm}^{2}(\rho)$ given by Eqs. (214) and (215) can be regarded as two effective potentials, and the fact of $\omega_{+}^{2}(\rho)\neq\omega_{-}^{2}(\rho)$ implies that photons with different helicities have different probabilities of escaping from a Schwarzschild black hole. In a word, once the gravitational spin-orbit coupling is taken into account, the original radius of the circular photon orbit will split into two different radiuses corresponding to photons with the helicities of $\pm 1$, respectively, which is also related to the fact that the spin-orbit coupling interaction induces a splitting of energy levels. As a result, on a given orbit, photons with the same wave number vector but different helicities, would have different probabilities of escaping from a Schwarzschild black hole. This may imply that photons from a Hawking radiation are partially polarized, rather than completely disorder.

## 8. Discussions and Conclusions

In Ref. [1] all spinors equivalent to tensors are taken as two-component forms, in our formalism, the $(1,0)\oplus(0,1)$ spinor equivalent to the electromagnetic field tensor is a six-component form, which is more convenient and straightforward. The typical significance of the $(1,0)\oplus(0,1)$ spinor description of the photon field lies in the research of the spin transport of photons. When we only concern the quantum-mechanical behaviors of single photons, it is more convenient for us to describe the photon field in terms of the $(1,0)\oplus(0,1)$ spinor. In particular, it is advantageous to study the spin-orbit interaction of the photon field based on the $(1,0)\oplus(0,1)$ spinor description, which is due to the fact that, it is impossible to construct a vector field for massless particles of helicity $\pm 1$ [2, 3], while the $(1,0)\oplus(0,1)$ spinor field exhibits simpler Lorentz transformation properties than (1/2, 1/2) four-vector field, and it can form the eigenstates of helicity $\pm 1$.

To describe the electromagnetic field in terms of the $(1,0)\oplus(0,1)$ spinor, we have discussed two equivalent representations of the $(1,0)\oplus(0,1)$ spinor, i.e., the chiral and standard representations, in a unified way. According to our study, the chiral and standard representations of the $(1,0)\oplus(0,1)$ spinor are respectively analogous to the chiral and standard representations of the Dirac field, but for the Pauli matrix vector $\boldsymbol{\sigma}=(\sigma_1,\sigma_2,\sigma_3)$ being replaced with the matrix vector $\boldsymbol{\tau}=(\tau_1,\tau_2,\tau_3)$. In other words, there are similar mathematical structures between the $(1/2,0)\oplus(0,1/2)$ spinor satisfying the Dirac equation and the $(1,0)\oplus(0,1)$ spinor satisfying the Dirac-like equation.

Starting from the $(1,0)\oplus(0,1)$ spinor description, the quantization theory of the photon field has been developed by means of a new approach, with many formulae being very useful to develop spin-photonics in our future work. Based on the $(1,0)\oplus(0,1)$ spinor formalism, some symmetries have been discussed from a new perspective and some new contents have been obtained. For example, the symmetry under the chiral transformation is studied in a unified manner, and two different interpretations of this symmetry are mentioned. Moreover, the symmetry under the operation of the chirality operator $\beta^5$ is discussed for the first time. In the past few years spin-orbit interaction (SOI) of light has been intensively studied in connection with the spin-Hall effect in inhomogeneous media and conversion of the angular momentum (AM) of light upon focusing and scattering, for them our work will provide a new research approach.

The spin-orbit interaction of the photon field in an inhomogeneous medium has been studied preliminarily and heuristically, where Eq. (138) can be regarded as optical analogues of the Dirac equation in the nonrelativistic limit. In our next work, we will present a thorough study on the spin-orbit interaction of photons based on this work, and try to lay a new foundation for spin-photonics, including the issues such as the photonic analog of the spin Hall effect, etc.

By means of the $(1, 0)\oplus(0, 1)$ spinor description of the photon field, one can treat the photon field in curved spacetime by means of spin connection and the tetrad formalism, which is of great advantage to study the gravitational spin-orbit coupling of photons. We have studied the effect of gravitational spin-orbit coupling on the circular photon orbit in the Schwarzschild geometry, and have reached a conclusion that, once the gravitational spin-orbit coupling is taken into account, the original radius of the circular photon orbit will split into two different radiuses corresponding to photons with the helicities of $\pm 1$, respectively, which is also related to the fact that the spin-orbit coupling interaction induces a splitting of energy levels. As a result, on a given orbit, photons with the same wave number vector but different helicities, would have different probabilities of escaping from a Schwarzschild black hole, which may imply that photons from a Hawking radiation are partially polarized, rather than completely disorder.

It is not necessary for the $(1,0)\oplus(0,1)$ description to be the photon field, provided that there is another massless vector field (e.g., they interact with matter just via the gravitational field). Moreover, there may be a chiral vector field described by Eq. (109).